\documentclass[letterpaper]{article} 
\usepackage{aaai25}  
\usepackage{times}  
\usepackage{helvet}  
\usepackage{courier}  
\usepackage[hyphens]{url}  
\usepackage{graphicx} 
\usepackage{natbib}  
\usepackage{caption} 
\usepackage{algorithm}
\usepackage{algorithmic}
\usepackage{xspace}

\usepackage{tcolorbox}
\usepackage{booktabs}

\usepackage{enumitem}
\usepackage{xcolor}
\usepackage{hyperref}

\usepackage{longtable}
\usepackage{url}

\usepackage{newfloat}
\usepackage{listings}
\DeclareCaptionStyle{ruled}{labelfont=normalfont,labelsep=colon,strut=off} 
\floatstyle{ruled}
\newfloat{listing}{tb}{lst}{}
\floatname{listing}{Listing}
\title{RelAItionship Building: Analyzing Recruitment Strategies for Participatory AI}
\author {
    Eugene Kim,
    Vaibhav Balloli,
    Berelian Karimian,
    Elizabeth Bondi-Kelly\equalcontrib, 
    Benjamin Fish\equalcontrib %
}
\affiliations {
    University of Michigan\\
    genakim@umich.edu, ecbk@umich.edu
}

\usepackage{bibentry}

\begin{document}

\maketitle

\begin{abstract}
Participatory AI, in which impacted community members and other stakeholders are involved in the design and development of AI systems, holds promise as a way to ensure AI is developed to meet their needs and reflect their values. However, the process of identifying, reaching out, and engaging with all relevant stakeholder groups, which we refer to as recruitment methodology, is still a practical challenge in AI projects striving to adopt participatory practices. In this paper, we investigate the challenges that researchers face when designing and executing recruitment methodology for Participatory AI projects, and the implications of current recruitment practice for Participatory AI. First, we describe the recruitment methodologies used in AI projects using a corpus of 37 projects to capture the diversity of practices in the field and perform an initial analysis on the documentation of recruitment practices, as well as specific strategies that researchers use to meet goals of equity and empowerment. To complement this analysis, we interview five AI researchers to learn about the outcomes of recruitment methodologies. We find that these outcomes are shaped by structural conditions of their work, researchers' own goals and expectations, and the relationships built from the recruitment methodology and subsequent collaboration. Based on these analyses, we provide recommendations for designing and executing relationship-forward recruitment methods, as well as reflexive recruitment documentation practices for Participatory AI researchers.

\end{abstract}

\section{Introduction}

Artificial intelligence for social impact (AISI) is being developed and deployed in many domains with promise to  address pressing societal challenges,
including healthcare \cite{kamineni_prospective_2023} and conservation \cite{tuia_perspectives_2022}. 
However, without involving stakeholders that are potentially affected by AI, there are risks to deploying AI broadly, such as encoding racism into AI systems \cite{benjamin_race_2019, obermeyer_dissecting_2019} or supporting the centralization of power \cite{birhane2022values}. 
Using participatory methods to carefully engage with stakeholders and impacted community members \textit{while} developing AI systems realizes values of collective governance and socially beneficial AI \cite{bondi_envisioning_2021, birhane_power_2022, costanza-chock_design_2020}. A body of work has been referring to the incorporation of participatory methods, drawn from theories of participatory design in computing and beyond, into AI development as ``Participatory AI'' \cite{kulynych_participatory_2020}.

This ethos of participation, while shared by many who study or aim to implement AISI and Participatory AI, can be difficult to achieve in practice \cite{lin_come_2024}. Barriers to implementing Participatory AI include a lack of clear standards and best practices \cite{delgado_participatory_2023}. Also, while Participatory AI idealizes the inclusion of marginalized  and impacted communities, 
AI practitioners report difficulties in recruiting and engaging with certain impacted communities \cite{cooper_fitting_2024, kallina_stakeholder_2024}. Despite the crucial impact of recruitment on who participates \cite{vines_configuring_2013}, there is no systematic practical guidance on  recruitment methods that AI researchers can use, especially for those who are new to participatory methods.

 We aim to develop practical guidance on recruitment strategies by studying the current practice of AI researchers. Our main research questions in this work are: 1) What are the recruitment strategies used by AI researchers? 2) How do methods impact goals of Participatory AI, and 3) How can we improve outcomes of recruitment methods? First, we review a corpus of 37 AI projects to characterize and describe the current recruitment strategies used in AI, and note that recruitment methods are often not documented sufficiently for purposes of reproducibility and transparency. Then, we interview five AI researchers and practitioners to learn what methods tend to be more successful and why. By analyzing both our corpus data and the interviews, we learn that structural conditions, researchers' own goals and expectations for the project, and relationships between researchers and community groups are key factors that influence the outcomes and the kinds of impact accomplished by the AI system (Table \ref{table:questions_takeaways}). We then introduce recommendations to improve the outcomes of recruitment methodology towards achieving Participatory AI.

In the rest of the paper,  we will provide an overview of related work in human-centered design and Participatory AI, discuss the methods we used for our review of existing practices of participation and our semi-structured interviews, the results of our analyses, and then introduce our recommendations for the design of recruitment strategies.

\begin{table*}[!t]
\centering
\begin{tabular}{|p{0.4\linewidth}|p{0.5\linewidth}|}
\hline
\textbf{Questions \& Data} & \textbf{Takeaways} \\
\hline
What are the recruitment strategies used in AI? \newline \textbf{Data:} Corpus &
\begin{itemize}
    \item Decision users and data creators are often the ones recruited.
    \item Researchers typically initiate recruitment.
    \item Most recruitment is done via organizations and social media, websites, advertisements, and formal events.
    \item Many details are not documented.
\end{itemize} \\
\hline
How do methods impact goals of Participatory AI? \newline \textbf{Data:} Interviews &
\begin{itemize}
    \item Researchers' capacities to engage in participatory projects are limited by structural conditions.
    \item Researchers' expectations and goals shape the methods they use and the relationships built with stakeholders.
    \item Project outcomes depend on the relationships built by the researchers and participants.
\end{itemize} \\
\hline
How can we improve outcomes of recruitment methods? \newline \textbf{Data:} Corpus, Interviews &
\begin{itemize}
    \item Relationship building is crucial when designing recruitment methodologies.
    \item Building more processes and structural support, such as institutions that facilitate connections, can help improve recruitment methods.
    \item Recruitment methodologies should be documented.
\end{itemize} \\
\hline
\end{tabular}
\caption{Research Questions and Takeaways}
\label{table:questions_takeaways}
\end{table*}

\section{Related Work}\label{sec:related_work}

\subsection{Participatory Methods in Computing}

Participation as a research methodology in computing draws from the fields of participatory design \cite{spinuzzi_methodology_nodate, asaro_transforming_2000} and participatory action research \cite{costanza-chock_design_2020}.
Participatory approaches' focus on participant knowledge and deconstructing knowledge hierarchies \cite{frauenberger_pursuit_2015} lends itself well to design methodologies that aim to achieve socially just outcomes by elevating marginalized voices. Because of this, participatory approaches are key components to design methodologies in computing oriented towards social justice \cite{abebe_anti-racist_2022, irani_postcolonial_2010, bardzell_towards_2011}, and calls to apply principles from participatory research traditions to AI are increasing \cite{klein_data_2024, costanza-chock_introduction_2020, mohamed_decolonial_2020}. 

At the same time, participatory methods in AI development have been critiqued for ``participation washing," in which participatory methods benefit the designers at the expense of the participants and manufacture consent for decisions made undemocratically \cite{sloane_participation_2022}. In response to these critiques, scholars have worked to empirically characterize current Participatory AI practice \cite{cooper_fitting_2024,groves_going_2023, kallina2025stakeholder}, develop conceptual frameworks to evaluate existing participatory projects in AI according to the goals of participant empowerment and democratizing AI governance \cite{birhane_power_2022,sloane_participation_2022,berditchevskaia_participatory_2021,feffer_preference_2023,bondi_envisioning_2021, corbett_power_2023, kallina2025stakeholder}, and provide high level guidance and values to design participatory methods to achieve these goals \cite{berditchevskaia_participatory_2021,feffer_preference_2023,delgado_participatory_2023,gilman_democratizing_2023,bondi_envisioning_2021}. 
The critical questions raised by previous work indicate that both practice and the social context of participatory methods are important in determining whether Participatory AI results in equitable, democratic AI development and systems. 

\subsection{Recruitment in Participatory AI}
A key concern in Participatory AI is studying how technology designers shape the participatory process. \citet{cooper_fitting_2024} examine the practices that designers, dubbed ``participation brokers," use to shape who participates and how they participate in AI design. They note the degree of power that AI researchers have in determining these practices, and observe that AI researchers frequently form partnerships with community groups  \cite{cooper_fitting_2024}. In another vein, \citet{lin_come_2024} studies how AISI partnerships often fail to meet the needs of community partners due to issues such as funding constraints and mismatched  expectations of AI's capabilities. They suggest that AI designers follow the lead of community members early in the project for AISI projects \cite{lin_come_2024}. Power and values-based recommendations for designing participatory methods are other key points  \cite{costanza-chock_design_2020, harrington_deconstructing_2019}.

Yet, there is still a need to provide practical guidance for participatory methods in AI. \citet{kallina_stakeholder_2024} interview AI practitioners in corporate and organizational contexts, finding that 70\% of respondents reported a lack of inclusivity in their samples  \cite{kallina_stakeholder_2024}. Respondents also note that certain stakeholder groups are difficult to identify and contact, in addition to having low incentives to participate. Other AI practitioners in corporate contexts similarly report needing clearer guidance as well as incentive structures to put participatory methods into practice \cite{delgado_participatory_2023,groves_going_2023, kallina2025stakeholder}. As these works note, recruitment is key for achieving goals of Participatory AI; it configures who participates in the first place, and can impact the goals of sharing design control with impacted communities \cite{vines_configuring_2013} and the values embedded in and performance of machine learning models \cite{diaz2024makes}. We aim to contribute to this body of work by clarifying the practice of participatory methods, particularly for recruitment.

\begin{table*}
\begin{tabular}{|p{0.1\linewidth}|p{0.4\linewidth}|p{0.4\linewidth}|}
\hline
 & Diverse and representative samples & Equitable relationships with communities \\ \hline
  Computing &  Oversample minoritized groups, work with communities, remove barriers to participation, build participant trust, and use inclusive demographic qualities \cite{roscoe_designing_2021} & Work with community groups in developing recruitment strategies \cite{university_of_michigan_stpp_program_and_detroit_disability_power_and_the_detroit_justice_center_community_2024} \\ \hline
  Civics &  Email and postal mail \cite{sharp_online_2010} & Reach out to grassroots organizations, community leaders and local media \cite{nabatchi_democracy_2012} \\ \hline
  Social Sciences & Participate in existing communities to gain access, snowball sample,
  consider participants' perception of research \cite{hollingshead_research_2012} & Work with advocates and leaders of community groups \cite{obrien_rigor_2022} \\ \hline
  Community Engagement & Develop culturally-specific engagement materials,  work with community liaisons \cite{mancera_utilizing_2022} & Work with advocates and leaders of community groups, provide compensation \cite{beyond_fixed} \\ \hline
  Health and Medicine & Call and text using phone numbers from EHR \cite{langer_recruitment_2021} & Offer pay and accommodate meeting needs, and recruit through organizations, social/print media, clinics, snowball sampling  \cite{leslie_recruitment_2019} \\ \hline

\end{tabular}
\caption{Non-exhaustive set of recruitment strategies in several fields}
\label{table:related}
\end{table*}

\subsection{Recruitment Methods} \label{sec:recruit}

Existing recruitment methodologies seek to accomplish primary research goals, such as conducting surveys and interviews \cite{heerman_recruitment_2017}, participant observation and other ethnographic methods \cite{negrin_successful_2022}, co-design practices, and more. Other methodologies, drawn from research traditions such as participatory action research, also ensure diverse and representative sampling, encouraging equitable relationships with communities, and building organizational capacity \cite{university_of_michigan_stpp_program_and_detroit_disability_power_and_the_detroit_justice_center_community_2024,obrien_rigor_2022, beyond_fixed}. We refer to Table \ref{table:related} for a brief overview of existing recruitment methods for a variety of purposes.  
Social scientists interviewing online communities, for example, highlight strategies such as participating in existing communities, using snowball sampling, and considering whether participants want to participate in the research projects \cite{hollingshead_research_2012}. This suggests approaches to recruitment that are relational and dependent on the researcher's positionality.

Reflection on 
recruitment in the context of researchers' specific positionalities and power imbalances is also prevalent. For example, relationships and institutional context impact recruitment, the research questions  that are pursued, and participant trust in researchers \cite{le_dantec_strangers_2015,small_qualitative_2022,gibson_racial_2003}. In particular, \citet{sum__2025} examine how exploitative relationships between the university and local communities influence  researchers' partnerships with community-based organizations. They ask researchers to build sustained relationships with community members, adopt individual tactics to support communities, and encourage institutions to revest in communities \cite{sum__2025}. Similarly, recruitment methods for non-``WEIRD" populations must also be contextual, flexible, and focus on needs and existing technological practices to be successful \cite{de2025gets}.

In short, much recruitment literature  exists that AI researchers can draw on.  Yet, little work has studied recruitment methods in AI 
or how those methods impact  Participatory AI goals. We aim to help fill these gaps, towards improving recruitment methodology for Participatory AI.

\section{Methods}

We investigated recruitment practices in projects that use participatory methods during the design and development of AI by: 1) creating and analyzing a corpus of 37 projects in this space; 2) conducting institutional review board-approved semi-structured interviews with 5 AI researchers.

\subsection{Corpus Creation and Analysis} \subsubsection{Assembling the Corpus}
We conducted a literature review in order to build a corpus of AI projects that represented a diversity of modes of participation, as defined by \citet{sloane_participation_2022}. 
We drew case studies from existing reviews of Participatory AI, particularly  \cite{birhane_power_2022,delgado_participatory_2023,feffer_preference_2023,berditchevskaia_participatory_2021,kulynych_participatory_2020}, case studies of AI for Social Impact \cite{tambe_ai_2022},  
recent publications at venues like FAccT, EAAMO, CHI, and AIES, and searches on Google Scholar for related terms like ``Participatory AI.'' 
We adopted this approach 
in an effort to include projects using a wide variety of participatory methods not explicitly labeled as such, including projects where participation served a variety of goals (e.g.\ participation as labor vs.\  consultation \cite{sloane_participation_2022}) and  impacts (e.g.\  participant  empowerment  \cite{delgado_participatory_2023}). Appendix\footnote{\url{https://realize-lab.github.io/participaite}} Tables \ref{tab:country}, \ref{tab:domains}, and \ref{tab:entry_point} provide an overview of the countries represented by our sampled projects, the domains they were in, and the participation entry point (as defined by \cite{corbett_power_2023}), respectively. To summarize, our corpus includes a wide variety of  domains and both Global South and Global North contexts, with a majority of projects coming from Global North contexts. In addition, the majority of participation took place during ``dataset development" and ``deployment and monitoring" entry points. 
We determined saturation was reached and concluded the literature review when novel recruitment categories in AI projects were no longer identified.

\subsubsection{Analyzing the corpus} After compiling the corpus, we used inductive coding \cite{braun_using_2006} with a goal to explore high-level themes and categories to characterize the recruitment methodologies used by AI researchers and practitioners. We found broad categories of recruitment methods, stakeholder types, and the actors who initiated the recruitment. We also coded each project by the level of empowerment afforded to the participants in the project \cite{delgado_participatory_2023}, to help us explore the question of how recruitment methods impact goals of Participatory AI. Two authors (EK, BK) coded the papers. Due to the small number of coders, we focused on building coding consensus via weekly meetings. 
Results are in Section~\ref{sec:corpus}.

\subsection{ParticipAIte}

We included the high-level categorizations from our qualitative coding into a 
preliminary database, which we call ParticipAIte\footnote{\url{https://realize-lab.github.io/participaite}}, and populated it with papers from our corpus. It includes information on recruitment methodology, including: a list of involved stakeholders, the recruitment strategy, who initiated recruitment, the specific call for participation, and  lessons learned (see Appendix \ref{sec:participaite}). 

\subsection{Semi-Structured Interviews}

For the semi-structured interviews, we aimed to gather more detail about the recruitment methodologies surfaced in our corpus, particularly from  
AI researchers with prior experience in recruiting participants for AI projects. 
Follow-up studies should also incorporate community group perspectives, which are essential for a more complete analysis of the empowerment of community groups.

\begin{table*}
\begin{tabular}{|l|l|l|l|l|}
\hline
ID & Role & Project experience discussed &  Context & Academic Training \\ \hline
P1 & Academia & AI for humanitarian purposes & GN & Computer Science \\ \hline
P2 & Academia & AI for environmental and humanitarian purposes &  GS \& GN & Computer Science \\ \hline
P3 & Academia & AI for humanitarian purposes  &  GS \& GN & Computer Science \\ \hline
P4 & nonprofit & AI and tech policy focused on marginalized groups' concerns  &  GN & Information Sciences \\ \hline
P5 & Academia & AI for humanitarian purposes  &  GN & Computer Science \\ \hline
\end{tabular}
\caption{Participants and descriptions (including Global South (GS) and/or Global North (GN) context focus)}
\label{table:participant_descriptions}
\end{table*}

We contacted ten researchers that were known to us to be researchers with Participatory AI experience 
via emails and direct messages, and interviewed five of them. Our selection was prioritized by our goal to converse with researchers who may have varied experiences in recruitment methodology, including the domain, format of participation, and goal of the participatory process. We acknowledge that the selection and response of participants were biased by our existing personal networks, interests, and reputations within certain research communities. See  Table \ref{table:participant_descriptions} for an overview of participant backgrounds.

Interviews lasted an hour, and were conducted via the video conferencing program Zoom. All interviews were recorded and transcribed with interviewee permission. We provided \$50 gift cards as compensation. The study was determined by the University of Michigan Institutional Review Board to be exempt from ongoing review (HUM00266324).

Interviews had two main components. For both components, participants had access to the ParticipAIte database, which contained an entry of one of their existing projects, as well as a  project in our corpus. First, we asked about interviewees' previous and current experience with recruitment methodology, using database entries describing their recruitment methodology that we assembled from existing work about the projects. We aimed to elicit detailed narratives behind their experiences of developing recruitment methodology.
The second main component was to discuss another recruitment case study in order to discuss documentation practices that would help with reproducibility. Finally, we asked for open ended, unstructured feedback on documentation about recruitment. See Appendix \ref{sec:interview_questions} for full interview questions.

As with the corpus, we performed an inductive coding on the interviews. Following thematic analysis \cite{braun_using_2006}, we developed higher level themes by analyzing data in context with the entire dataset, as well as descriptive themes. The results are in Sections~\ref{sec:results} and \ref{sec:lessons}.

\section{Recruitment Methodologies}\label{sec:corpus}

Our corpus analysis was based on descriptive categories of recruitment methodologies, including: \textit{Who is recruited}, \textit{who recruits},  \textit{recruitment strategy}, and 
\textit{recruitment impacts on participant empowerment}. 
We also noted  patterns in documentation. The Appendix lists the full corpus and analysis.

\subsubsection*{Who is recruited?}
\begin{figure}
  \centering
  \includegraphics[width=0.75\columnwidth]{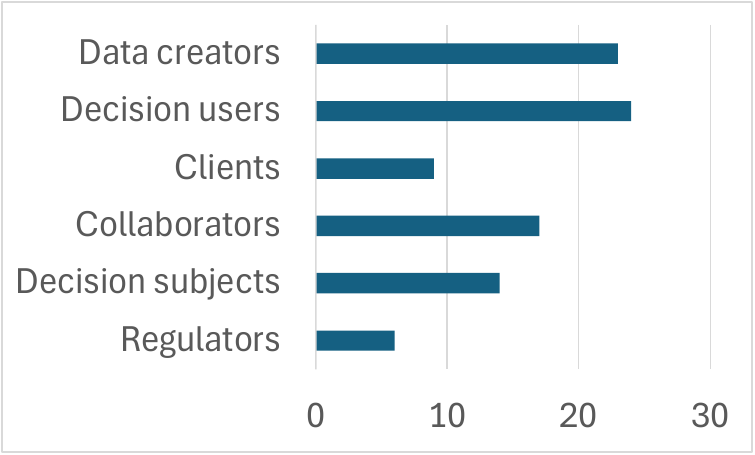}
  \caption{``Those recruited'' papers (Appendix Table \ref{table:stakeholder_groups})}
  \label{fig:who_gets_recruited}
\end{figure}
In general, projects in our corpus tended to recruit either end users of an AI system (``Decision users") or data labellers and workers (``Data creators") as participants. Results are presented in Fig.\ref{fig:who_gets_recruited} and Table \ref{table:stakeholder_groups} in the Appendix. 

Works in the corpus also sometimes detailed the relevant stakeholder groups that they did not aim to recruit from \cite{lee_webuildai_2019,yu_keeping_2020,newman_kudu_nodate,cheng_soliciting_2021,sendak__2020}, acknowledging how recruitment strategies determine who can or cannot participate. 
Some projects excluded stakeholders due to the challenges  of engaging said stakeholders and their more indirect relationships to the system being developed, such as patients in a tool designed for healthcare workers \cite{sendak__2020}. 

In our corpus, authors also noted  that recruitment can shape barriers to participation.    
For example, enabling online participation and flexible scheduling were some strategies to ensure equitable participation and broader recruitment 
\cite{queerinai_queer_2023,kuo_understanding_2023,lee_webuildai_2019}. A lack of inclusion  due to barriers could in turn lead to bias and generalization challenges \cite{awad_moral_2018,calacci_bargaining_2022,marian_algorithmic_2023,bakker_fine-tuning_2022}.

\subsubsection{Who recruits?}
\begin{figure}
  \centering
  \includegraphics[width=0.75\columnwidth]{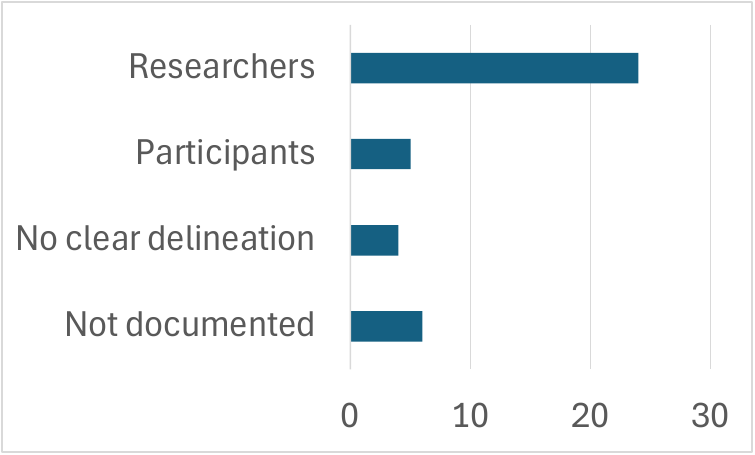}
  \caption{``Who recruits'' papers  (Appendix Table \ref{table:who_initiated})}
  \label{fig:who_recruits}
\end{figure}

In addition to differences in  who was recruited, there were also different stakeholders doing the recruitment (see Appendix Table \ref{table:who_initiated} and Fig.  \ref{fig:who_recruits}). AI researchers or practitioners most frequently conducted recruitment in our corpus. Furthermore, some of our corpus provided descriptions of how the researchers or practitioners' previous experiences with the communities of their participants allowed access to  research sites \cite{lee_webuildai_2019,wilder_clinical_2021,eitzel_assessing_2021}. Our corpus also had a few rich descriptions of collaboration building between researchers and community partners \cite{delgado_uncommon_2022,service_goat_2019}. 

There were a few instances where the line between researcher and participant was blurred, as both parties engaged in the process of research \cite{queerinai_queer_2023,nekoto_participatory_2020}.
In our corpus, we also found several examples of participant-led research, where researchers were solicited by communities or developed the research problem in collaboration with communities (see Appendix for a  list). 

\subsubsection*{Recruitment Strategy}

\begin{figure}
  \centering
  \includegraphics[width=0.75\columnwidth]{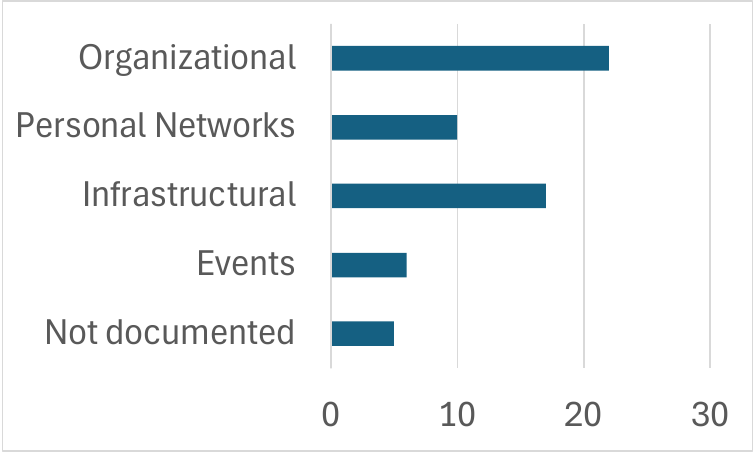}
  \caption{``Recruitment strategy'' papers  (Appendix Table \ref{table:recruiting_strategy})}
  \label{fig:recruitment strategy}
\end{figure}

Recruitment strategies were also quite varied, as shown in Fig. \ref{fig:recruitment strategy} and Table \ref{table:recruiting_strategy} in the Appendix. Recruitment strategies included working with organizations (organizational), drawing on personal networks,  using social media or other mass media (infrastructural), and attending events like conferences (Table~\ref{table:recruiting_strategy}  details specific strategies in these categories). We found that most recruitment in our corpus was based on connecting to organizations, for example, by reaching out to community advocacy organizations or nonprofit organizations.

\paragraph{Empowerment} Our corpus showcased several ways for recruitment methods to meet goals of equity and empowerment. For example,  researchers ensured that their methods followed community norms via conversations with the community \cite{zhu_value-sensitive_2018}, relevant domain experts \cite{kuo_understanding_2023}, and local institutions \cite{newman_kudu_nodate}. Works in our corpus also aimed to build trust with marginalized communities by recruiting institutional actors that worked with these communities \cite{wilder_clinical_2021,kuo_understanding_2023,young_toward_2019,frauenberger_pursuit_2015} and key organizers of  communities 
\cite{calacci_bargaining_2022,marian_algorithmic_2023}. 

While crowdsourcing has been noted as a form of participation that  is considered labor,  is more at the level of consulting or including (using \citet{delgado_participatory_2023}'s scale of participant empowerment), or offers a low amount of agency  to  participants  \cite{nekoto_participatory_2020}, 
careful crowdsourcing can also empower community groups. A prominent example is Digital Workerism \cite{calacci_organizing_2022}, where crowdsourcing was used to provide data for a worker-led campaign to investigate the impact of algorithmically determined wages on working conditions. 
In our corpus, researchers and practitioners used intentional strategies to reach marginalized communities during crowdsourcing, including decentralized and internet-based recruitment strategies to reach community members over large geographical divides \cite{queerinai_queer_2023,nekoto_participatory_2020}, requiring few resources to recruit and participate \cite{li_participation_2023,queerinai_queer_2023}, and reaching out to strategic allies during the recruitment process \cite{nekoto_participatory_2020,irani_turkopticon_2013}. Offering some direct benefits to participants, such as easing existing workflow challenges \cite{kelling_ebird_2012}, or ensuring that the resulting data and knowledge production were of immediate benefit to participants \cite{li_participation_2023}, were also important practices.

\subsection{Documentation of Recruitment Strategies}
So far, we have focused on those parts of recruitment methodologies that are documented in our corpus.  However, projects varied in level of documentation. For example, Figs. \ref{fig:who_recruits} - \ref{fig:recruitment strategy} showed multiple projects lacking enough documentation for us to determine a categorization. Some projects described the specific actors and institutions they work with, while others only referred to the kind of institution. 
While omitting specific names to protect privacy is certainly an important consideration, we believe documentation of other aspects of recruitment are important for purposes like reproducibility and transparency. For instance, we noted several helpful examples where authors detailed    
reasons for excluding or including certain stakeholders, how researchers leveraged existing personal relationships during recruitment, or specific motivations for selecting certain recruitment strategies. These were not documented across the entire corpus. 
As such, we were motivated to interview AI researchers and learn about their experiences with recruitment methodology.

\section{Results from Interviews} \label{sec:results}

Our interviews focused on interviewees' previous experience with recruitment. To briefly summarize, each of the interviewees identified as researchers engaging in recruitment. 
They were primarily  concerned with recruiting collaborators, mostly community organizations, other grassroots advocacy and activist groups, and members of specific communities, for co-design projects. 
To meet these recruitment goals, interviewees largely began by identifying partners with experience and existing relationships with communities the interviewee wished to engage with as a starting point, and reaching out most frequently  through personal networks and cold calling.  
Once they established a relationship, e.g., with a nonprofit or government agency, they then often  partnered with  organizations for many core tasks, including: surveying  communities, recommending policy, and introducing  interviewees to other stakeholders. This aligns with our findings from the corpus, which  showed the prevalence of   organizational and personal network recruitment (Fig. \ref{fig:recruitment strategy}), and slightly differs from formal social science recruitment methods, such as surveying or random sampling. 

We now analyze and present our detailed findings in three connected themes. First, we explore how researchers' capacities to begin and engage in participatory projects are inherently limited by structural conditions. Second, we explore how researchers' expectations and goals shape the methods they use, and how methods can impact the relationships built with stakeholders in the domain. Finally, we argue that outcomes of participatory projects are fundamentally based on the relationships built by the researchers and participants.

\subsection{Structural Challenges to Recruitment}\label{sec:participatory_challenges}

Before we begin to address the recruitment methodologies that are used by researchers, as well as their outcomes, we need to discuss what constraints and limitations researchers face. 

\paragraph{Funding} A first barrier is funding. Researchers noted: \textit{``to support this kind of project is not as easy compared to say, defense and security,"} pointing to a clear lack of funding support for projects supporting community-based organizations (P2). Not only do the researchers need to get funding for themselves, they also need to provide funding for their collaborators (P2). Funding for the participating community-based organization is particularly important, as participants, staff, and the organization need to be paid for their time and any additional coordinating labor (P4). One interviewee also addressed the fact that continued funding is a significant problem to achieving participatory projects with a long-term goal and focus: \textit{``there's not a ton of funding available for  maintenance of systems"} (P3).

\paragraph{Time and effort} A second barrier is the time commitment added on from deciding to pursue a participatory project. For example, researchers discussed how participation from organizations and communities required senior researchers to \textit{``identify a problem at the right scope"} (P2) for a student to work on, as well as prototyping and creating pilot experiments to determine operational feasibility (P5). Due to the considerable time investments required from both the AI practitioners and the involved stakeholders, both parties would need to commit seriously to the length of the project (P2).

\paragraph{Logistics and coordination}  These challenges are compounded by logistical and legal concerns.  P5, for example, was hampered by such issues when attempting to include a certain stakeholder group. Coordinating with communities and stakeholder groups can require considerable investment and project management, including overhead not necessary when working alone, such as additional meetings (P5), testing prototypes in the field (P5), and engaging in the domain (P2).

\paragraph{Ethical engagement and building trust} The entrenched power dynamics between researchers and community groups is another barrier. One researcher discussed the difficulty in gaining the trust of certain groups: \textit{``you're always  working in the wake of whatever the university that you work at has done to that community"} (P4). Cognizance of structural harms and of ways that marginalized groups are often further marginalized by exploitative projects means that researchers need to evaluate whether the community groups and organizations that they are working with even have the capacity to participate in these projects, and if participation is putting an undue burden on the community groups (P4). 

\paragraph{Lack of (relevant) expertise} Some researchers are not trained in engaging with and recruiting community members or groups. One researcher stated: \textit{``we just do not have any of the expertise needed to kind of navigate that world and do anything sensible"} (P1) when it comes to directly outreaching to community members. In fact, some interviewees
were not involved with  community engagement
or recruitment despite being involved in Participatory AI projects, as teams of researchers
had split into  community-engagement and  technical intervention subteams, with the technical researchers
relatively separated from decisions about engaging
with the community (P1, P5). 

Furthermore, technical expertise can also serve as a structural challenge in recruitment.  Several interviewees had explicit expectations to
develop specific technical contributions as a result of a collaboration due to career incentives.
This presents a very serious trade-off for interviewees:
\textit{``how much effort and time do you really want to
spend on these AI for social good projects that may not fully
leverage your technical expertise to work on?”}  They suggested that, for researchers interested in creating social impact without engaging in the additional labor required to actively involve community organizations and groups,
foregoing recruitment and working on open-source projects
may be more efficient for achieving social impact (P2).

\subsection{Impacts of Challenges and Goals  on Recruitment}\label{sec:partner-search}

These challenges, coupled with researcher goals,  ultimately incentivized researchers to find partners as a means of finding participants. This, in turn,  shaped recruitment methods.

\subsubsection{Personal connections and partners}
Working with organizations proved to be an effective strategy for researchers to engage in participatory projects, particularly with organizations that had both the capacity to work with researchers and were already embedded in the communities of interest. \textit{This helped address challenges with logistics, engagement, expertise, and trust}, to name a few. 

For example, P4 noted that organizations with a grassroots orientation enable better connection to membership and an ability to directly bring community members into the process of Participatory AI, while P1, P2, and P5 stressed the importance of having a community partner that already had  connections to communities and the relevant expertise to work in these domains. P4 also found that going to the director of advocacy or programming at nonprofits was promising due to an indication that the organization had an increased capacity for collaboration.  These existing connections to partners with capacity were significant in starting a collaboration.

At the same time, the structural barriers that made partnering with community groups  more effective also led to challenges. 
P2 summarized the challenge as \textit{``I think in general, it's not easy, because we have limited capacity, and these nonprofits also have their own agenda and priorities"} (P2).  Furthermore,  one interviewee reflected how success is often determined by an intuition surrounding the quality of the relationship: \textit{``I do feel like a lot of recruitment happens in sort of...a ground-up way...a lot of it is vibes"} (P3). Ultimately, \textit{``finding [any] collaborator is easier than finding a good collaborator"} (P2). 
 
All of our interviewees reported heavily relying on personal networks in their recruitment processes; including academic and personal networks.  
In fact, our interviewees reported that relying on personal networks led to successful academic outcomes (P1-P5).  

\paragraph{Mixed success from cold calls}
To find partners, the alternative to leaning on connections was cold calling potential partners.   Cold calls, whether through email, phone, or direct messages, were reported to have mixed success rates - failures of collaborations starting from cold calls included no responses (P2, P4), or collaborations that were short-term or underwhelming (P2). P2 shared an anecdote where a close colleague emailed dozens of nonprofits in their local area, but were unable to successfully create a collaboration.  While researchers admitted that they could not always know why a cold-call attempt failed, given that people are very busy and have other priorities, refusal to participate could indicate a deeper mistrust of researchers and the institutions they represent (P4), adding yet another difficulty to the process.

Establishing a relationship from cold calls is often easier when the researcher can lean on their existing reputation in the community (P2, P3). P4 said that \textit{``If someone's not familiar with you, I think that some of this work comes across like weird marketing research ... like getting a cold call from an organization that wants to pay you..."} (P4). To address this challenge and to successfully get a response, interviewees suggested showing examples of previous existing projects (P2), emphasizing compensation (P4), and making a case that the community group  has novel, important contributions to make (P4).

\paragraph{Challenges to achieve representativeness}
All of our interviewees engaged with goals of representativeness when they decided how and who to recruit. 
For instance, researchers were motivated to use cold calls to bring in partners they otherwise would not. P4 reflected: \textit{``It would be a lot easier if
we did only rely on our personal networks, because the [organizations]
would be much more likely to participate”}, as they may already see the value in the project, but 
\textit{``if you stay too close to home, then you're just talking to the same people with the same interests. So often, it's cold calls"} (P4). 

One key challenge is that capacity limits interviewees' ability to achieve ideals of representativeness. Even though our interviewees aspired to include all relevant groups, this was not practically possible. 
One interviewee regretted not including one group of stakeholders, mentioning that even a small, biased sample would be better than having none  (P5). The reasons cited by researchers for not including these specific stakeholder groups were a lack of capacity to navigate the logistical (e.g., institutional review board) obstacles to include these groups, noting that doing so would have added potentially years to the research study (P5). 

Researchers also emphasized that  representativeness is a priority throughout project design and development; %
questions of how to make participation \textit{``successful and sustainable"} are not \textit{``specific to recruitment"} (P3). The way people participate and to what degree they can influence the system, for instance, are similarly important. Recruitment is key for achieving this goal of representativeness throughout a project.

\subsection{Relationship Impacts on Outcomes}
The challenges in  finding participants  underscore significant trade-offs between pursuing and building relationships with community groups, and ensuring participatory projects can meet researchers' career needs, such as milestones and publishing requirements.  
Given that these kinds of tensions are largely structural, we found that the interviewees responded  by trying to build longer-term relationships that were mutually beneficial to community partners and researchers.

\paragraph{Long-term partnerships}
Long-term partnerships were noted as having many benefits for interviewees and the partners they were working with. Ideally, when recruiting partners, researchers were looking for ``long-term collaborations" (P2). Long-term relationships were noted as valued by community groups, who have made it clear to researchers that parachuting into community to produce research and leaving is not acceptable (P2). Other benefits included producing several research papers over a stand-alone application (P2) and establishing new governance relations between designers, governors, and community organizations (P4).  Echoing concerns and critiques of exploitative projects,  
P4 described the frequently short-term nature of Participatory AI as something \textit{``arguably broken about Participatory AI."} 

Interviewees in our study discussed several more successful strategies for building longer-lasting, durable relationships between researchers and their collaborators (P1-P3). For example, P1 talked about a project in which the nonprofit center benefited, as the conducted research provided important administrative tasks and helped provide benefits to the community that the nonprofit served, despite the center choosing not to use the algorithm the researcher created in day-to-day usage. This relationship was mediated by a senior researcher's long-standing relationship with a collaborator in the domain who had years of experience collaborating with relevant nonprofits.  
In addition, P2 spoke to a model of relationship-building that was more scalable and accessible to junior researchers with fewer resources: senior researchers maintaining long-standing relationships with many partners and nonprofits to help scope projects for more junior researchers, and relying on existing long-term relationships between senior researchers and community partners. Thus, long-term relationships can be the basis of many separate projects, and provide continued impact for community partners.

\subsubsection{Relationship challenges and deployment}
Yet, long-term partnerships are difficult to achieve. One interviewee recalled a collaboration that broke down right before the deployment of an algorithm, despite the CEO of the nonprofit being very enthusiastic about the project in the initial stages. When reflecting on what could have been improved, the interviewee suggested that the researchers and the nonprofit needed an ``explicit collaboration" from the beginning to avoid failure: \textit{``Maybe just having them attend way more meetings, even on the [technical] side, would help them feel even more like part of this."} 
Unfortunately, in this case, the algorithm did not get to be deployed. Three of our interviewees (P1, P2, P5) reported that the algorithmic interventions  created for nonprofit partners were not still currently deployed after the researchers ``handed off" the project to the organizations. This happened when the project would end, and the researchers would no longer be involved in the project, often moving to other universities (P1 P2, P5), or when the nonprofit would fundamentally alter the way they collect data, rendering the algorithm obsolete (P1, P2). The challenges maintaining relationships between researchers and community-based organizations during the development of the algorithm  may have contributed to limited algorithmic impacts.

\paragraph{Outcomes beyond deployment} Sometimes, benefits to nonprofits or community partners existed beyond algorithmic impacts. Other benefits mentioned in our interviews
included receiving funding (P2, P4), gaining legitimacy in
the eyes of governments and other decision-makers (P3),
establishing data collection practices initiated by researchers
(P1), or even developing AI literacy (P2). This indicates, then, that
truly impactful participation might not be measured in terms
of algorithm deployment alone, but also holistically with respect to
other benefits.

\section{Discussion} \label{sec:lessons}

Given the challenges we found to successfully recruiting for Participatory AI, there is a clear need for improved recruitment methodology.  Design recommendations for Participatory AI from previous work can and should apply to recruitment methods, such as recommendations to examine and challenge power \cite{dignazio_data_2023}, to prioritize marginalized communities \cite{costanza-chock_design_2020}, and to ensure research meaningfully benefits the involved communities \cite{erete_i_2021}.   
The interviewees (see Section~\ref{sec:results}) wanted to achieve these kinds of goals, but the challenges they faced were significant. 
We now discuss possible approaches to improving recruitment strategies.

\subsection{Building Community}

Our findings highlighted the highly relational aspect of recruitment. From researchers' reputations to building trust to achieving representativeness, recruitment methodologies often succeeded or failed based on the relationships between researchers and communities.

We encourage researchers to expand conceptions
of what is a legible recruitment strategy and who
is able to use these strategies. 
For instance, in addition to finding partners, researchers can establish themselves within the networks of communities that they aim to work with by helping organize or attending workshops that focus on prominent issues in local communities, or regularly volunteering with or organizing alongside community groups. 
This echoes findings from studies for researchers to build personal relationships with the communities they work with \cite{sum__2025, harrington_deconstructing_2019}. Forefronting trust and building strong relationships in recruitment also helps tackle a core question -- who is actually participating in Participatory AI -- by encouraging participation from marginalized groups.

There are existing mechanisms for researchers to join community-building efforts. Interviewees mentioned building spaces for community groups and organizations to create relationships with AI practitioners, such as at the FAccT conference, where \textit{``if you make space for more and more and more community-based organizations and advocacy groups to present there, you do build real relationships"} (P4). Another interviewee (P2) mentioned creating academic spaces where a principal investigator can maintain many relationships with community organizations to support PhD projects, and summer programs where a senior academic matches undergraduate and masters students to community groups.  P3 discussed creating a formal network of community organizations and advocacy groups with similar concerns: \textit{``the project is more well known and global ... so sometimes people find us."} Maintaining such a network included holding regular events for interested groups and extending regular invitations to engage with the researchers, such as relevant research colloquiums and symposiums, resulting in long-term connections and collaborations over time (P3, P4).  Although labor-intensive, building spaces that can help connect expertise, potential funding, and need is a promising approach to growing personal and professional networks without an individual researcher or community needing to do it all themselves.

Once researchers are able to establish themselves within a network, it is far more likely that community organizations  will approach a researcher with a proposal. One interviewee said, after participating in a project with an organization that was organizing similar community groups, \textit{``this relationship unlocked all these relationships to local community-based organizations"} (P4).  Describing how these relationships eventually led to a collaboration, P4 described how \textit{``a contact was looking for some help from us . . . [and] as that conversation evolved . . . we realized there was a lot of overlap."}  Once an interviewee had one successful collaboration, they were able to focus on recruiting similar organizations with similar structural needs that could be addressed by the algorithmic intervention that they had created (P3), showing a reputational benefit for similar communities.

Besides benefits to the researcher, earlier engagement with communities can help all participants get involved at earlier stages of projects and better engage with other stakeholders, potentially building a sense of shared ownership and algorithmic interventions with a clear and sustained social impact. 
Moreover, technical interventions made with engagement from the relevant communities early on are key to anticipating needed changes in the design phase, reflecting recommendations to ``go out in the field" \cite{perrault_artificial_2020}. Earlier engagement also makes it easier to clearly agree on the potential benefits at the beginning of the project, using the findings in this work and others including \citet{lin_come_2024, ismail_public_2023, erete_towards_2025} as guidance to frame such conversations. When researchers are in community with the broader, local community, there are possibilities for directly recruiting community members themselves in a grassroots manner, rather than recruiting through formal institutions that represent them. 

While building community is promising to further Participatory AI goals, we acknowledge that the challenges detailed in Section~\ref{sec:results} remain. 
Individually, a researcher may need to adjust expectations and narratives to make the work \textit{``legible as a research contribution to a specific community"} (P3).   
Without changes to the institutions that support researchers and communities alike, fully overcoming many of these challenges, from funding to coordination to sustaining long-term relationships, will remain difficult. We believe that sharing challenges and successes will help improve  recruitment methodology.

\subsection{Reflexive Documentation}

During our analysis of the corpus, we saw significant disparities in the detail of documentation of recruitment strategies. When we interviewed researchers, we learned many more details about recruitment than from the researcher's published papers. These stories included the details and difficulties of finding funding, navigating specific social networks, engaging with specific actors to build relationships with community partners, and working with a set of institutions, such as universities and conferences, to help researchers build the relationships to recruit.  As such, we recommend better documenting recruitment practices.
Our recommendation to document these ``backstage" aspects \cite{bodker_participatory_2018} draws from a tradition of researcher reflexivity to improve transparency and create better recruitment methodologies in the field. Another benefit of introducing this reflexivity is to make the invisible work of recruitment and establishing relationships more visible to the research community, which can initiate change in how this labor is valued by institutions \cite{le_dantec_strangers_2015}. Reflexivity is also a tool that can allow researchers to more clearly understand the complexities of power dynamics that impact their research projects, including the hierarchies, interests, and worldviews driving the decisions in research \cite{miceli_documenting_2021}.
We now attend to three factors of reflexive documentation for participatory projects: social and personal networks, institutional support, and learning from refusal. These three factors are introduced as a starting point for reflexive documentation.

\paragraph{Personal networks}
In our corpus of Participatory AI projects, how recruitment leveraged personal networks  was not always very clear, yet our interviewees emphasized the importance of personal networks to their recruitment methodologies. We suggest that researchers record which relationships were key in building community partnerships, how these relationships were initially formed (as raised by P2), and how they resolved any conflicts in the relationships. Detailed narratives of recruitment are given in \citet{le_dantec_strangers_2015, delgado_uncommon_2022}, and \citet{tandon_hostile_2022}, but even brief descriptions would help provide additional context for others  recruiting.

\paragraph{Institutional support} 
As researchers navigate their own capacities for engaging in Participatory AI, guidance on finding funding is key. One interviewee in our study strongly suggested documenting how researchers recruited funding in Participatory AI projects (P2). Beyond sharing strategies for identifying such funding sources, it would be useful to share details about the structures of institutions that support Participatory AI development. As an example of this, one of our interviewees spoke to how replicating successful institutions in other universities that enable Participatory AI development could help their own university support similar endeavors to improve recruitment outcomes (P1).

\paragraph{Refusal and non-response}
Researchers tended to document the successful recruitment strategies, but often had stories of unsuccessful recruitment strategies that were not documented.   
Recruitment non-response and the failure of strategies to produce successful partnerships are also part of the recruitment process, and should be documented. Narratives of these failures and how they were overcome can provide starting points for other researchers designing recruitment methodology. As P5 put it, \textit{``finding out what went wrong is almost more useful than finding out what went right, because there are certain things that...everyone would do differently in each of the papers."}

Documenting the instances of refusal to participate can provide light on which communities are underrepresented in Participatory AI and why. AI designers have called for supporting refusal in design \cite{barocas2020not}, and P4 noted that \textit{``[academic and nonprofit spaces] can be... almost violent"}  for certain people, highlighting the importance of refusal as a mode of participation. Using data collected about refusal, Participatory AI can create methods to incorporate refusal as an active mechanism of governance and accountability, potentially building on models of informed refusal in bioethics \cite{benjamin_informed_2016}.

\subsection{Towards Representativeness}
Finally, representativeness and meaningful participation are key goals in Participatory AI, yet our interviews highlighted the difficulty achieving these goals due to logistical challenges and challenges with cold calling potential partners, for instance. 
Right now, finding partners and communities is done largely ad-hoc.  Even when done well, it is hard to learn what worked well, let alone to reproduce the methodology in new projects and new domains.  In part, this is a problem of documentation, but it is also because researchers may not have the same access or context other researchers had.  
We believe that there would be value in combining relationship building with a more systematic recruitment methodology. 
Ideally, a systematic recruitment methodology could be utilized by many researchers to improve representativeness, even in new contexts, and could allow repeated evaluation of methodology across many projects. 
A focus on recruitment methodology can highlight the key role that affected communities should play in developing AI.

\section{Limitations and Future Directions}

The selection of the corpus and interviewees reflect our positionalities as researchers based in a research institution in the Global North, as well as our personal networks and publication goals. 
In addition to interviewing AI researchers, community groups that have partnered with AI researchers could provide more insights on these challenges, particularly from their perspectives and needs. Existing recruitment practices of other kinds of practitioners of Participatory AI, such as community groups that organize and gather data about AI systems, are also under-represented and deserve further study. Together, these directions can help advance the goal of improving Participatory AI and AI for Social Impact practice. 

\section*{Acknowledgments}
We would like to thank our interviewees for taking the time to converse and share their perspectives with us. We would also like to thank the reviewers, as well as members of the Computational Social Science group, Realize Lab, the Political Economy and Algorithms group, Nathan Kim, Shreya Chowdhary, and others at the University of Michigan, for feedback at crucial stages of the paper. 

\bibliography{participaite-references-manual}

\begin{thebibliography}{99}
\providecommand{\natexlab}[1]{#1}

\bibitem[{Abebe et~al.(2022)Abebe, Amaryan, Beshai, {Ilene}, Gurgen, Ho, Hylton, Kim, Lee, Lewandowski, and {others}}]{abebe_anti-racist_2022}
Abebe, V.; Amaryan, G.; Beshai, M.; {Ilene}; Gurgen, A.~E.; Ho, W.; Hylton, N.~R.; Kim, D.; Lee, C.; Lewandowski, C.; and {others}. 2022.
\newblock Anti-{Racist} {HCI}: notes on an emerging critical technical practice.
\newblock In \emph{{CHI} {Conference} on {Human} {Factors} in {Computing} {Systems} {Extended} {Abstracts}}, 1--12.

\bibitem[{Asaro(2000)}]{asaro_transforming_2000}
Asaro, P.~M. 2000.
\newblock Transforming society by transforming technology: the science and politics of participatory design.
\newblock \emph{Accounting, Management and Information Technologies}, 10(4): 257--290.
\newblock Publisher: Elsevier.

\bibitem[{Awad et~al.(2018)Awad, Dsouza, Kim, Schulz, Henrich, Shariff, Bonnefon, and Rahwan}]{awad_moral_2018}
Awad, E.; Dsouza, S.; Kim, R.; Schulz, J.; Henrich, J.; Shariff, A.; Bonnefon, J.-F.; and Rahwan, I. 2018.
\newblock The moral machine experiment.
\newblock \emph{Nature}, 563(7729): 59--64.
\newblock Publisher: Nature Publishing Group.

\bibitem[{Bakker et~al.(2022)Bakker, Chadwick, Sheahan, Tessler, Campbell-Gillingham, Balaguer, McAleese, Glaese, Aslanides, Botvinick, and {others}}]{bakker_fine-tuning_2022}
Bakker, M.; Chadwick, M.; Sheahan, H.; Tessler, M.; Campbell-Gillingham, L.; Balaguer, J.; McAleese, N.; Glaese, A.; Aslanides, J.; Botvinick, M.; and {others}. 2022.
\newblock Fine-tuning language models to find agreement among humans with diverse preferences.
\newblock \emph{Advances in Neural Information Processing Systems}, 35: 38176--38189.

\bibitem[{Bardzell and Bardzell(2011)}]{bardzell_towards_2011}
Bardzell, S.; and Bardzell, J. 2011.
\newblock Towards a feminist {HCI} methodology: social science, feminism, and {HCI}.
\newblock In \emph{Proceedings of the {SIGCHI} conference on human factors in computing systems}, 675--684.

\bibitem[{Barocas et~al.(2020)Barocas, Biega, Fish, Niklas, and Stark}]{barocas2020not}
Barocas, S.; Biega, A.~J.; Fish, B.; Niklas, J.; and Stark, L. 2020.
\newblock When not to design, build, or deploy.
\newblock In \emph{Proceedings of the 2020 Conference on Fairness, Accountability, and Transparency}, 695--695.

\bibitem[{Benjamin(2016)}]{benjamin_informed_2016}
Benjamin, R. 2016.
\newblock Informed refusal: {Toward} a justice-based bioethics.
\newblock \emph{Science, Technology, \& Human Values}, 41(6): 967--990.
\newblock Publisher: SAGE Publications Sage CA: Los Angeles, CA.

\bibitem[{Benjamin(2019)}]{benjamin_race_2019}
Benjamin, R. 2019.
\newblock \emph{Race {After} {Technology}: {Abolitionist} {Tools} for the {New} {Jim} {Code}}.
\newblock Polity.
\newblock ISBN 978-1-5095-2639-0.

\bibitem[{Berditchevskaia, Peach, and Malliaraki(2021)}]{berditchevskaia_participatory_2021}
Berditchevskaia, A.; Peach, K.; and Malliaraki, E. 2021.
\newblock Participatory {AI} for humanitarian innovation.
\newblock \emph{London: Nesta}.

\bibitem[{{Beyond Chicago}(2020)}]{beyond_fixed}
{Beyond Chicago}. 2020.
\newblock Why Am I Always Being Researched?

\bibitem[{Birhane et~al.(2022{\natexlab{a}})Birhane, Isaac, Prabhakaran, Diaz, Elish, Gabriel, and Mohamed}]{birhane_power_2022}
Birhane, A.; Isaac, W.; Prabhakaran, V.; Diaz, M.; Elish, M.~C.; Gabriel, I.; and Mohamed, S. 2022{\natexlab{a}}.
\newblock Power to the {People}? {Opportunities} and {Challenges} for {Participatory} {AI}.
\newblock In \emph{Proceedings of the 2nd {ACM} {Conference} on {Equity} and {Access} in {Algorithms}, {Mechanisms}, and {Optimization}}, {EAAMO} '22, 1--8. New York, NY, USA: Association for Computing Machinery.
\newblock ISBN 978-1-4503-9477-2.

\bibitem[{Birhane et~al.(2022{\natexlab{b}})Birhane, Kalluri, Card, Agnew, Dotan, and Bao}]{birhane2022values}
Birhane, A.; Kalluri, P.; Card, D.; Agnew, W.; Dotan, R.; and Bao, M. 2022{\natexlab{b}}.
\newblock The values encoded in machine learning research.
\newblock In \emph{Proceedings of the 2022 ACM conference on fairness, accountability, and transparency}, 173--184.

\bibitem[{Bondi et~al.(2018)Bondi, Fang, Hamilton, Kar, Dmello, Choi, Hannaford, Iyer, Joppa, Tambe, and {others}}]{bondi_spot_2018}
Bondi, E.; Fang, F.; Hamilton, M.; Kar, D.; Dmello, D.; Choi, J.; Hannaford, R.; Iyer, A.; Joppa, L.; Tambe, M.; and {others}. 2018.
\newblock Spot poachers in action: {Augmenting} conservation drones with automatic detection in near real time.
\newblock In \emph{Proceedings of the {AAAI} {Conference} on {Artificial} {Intelligence}}, volume~32.
\newblock Issue: 1.

\bibitem[{Bondi et~al.(2021)Bondi, Xu, Acosta-Navas, and Killian}]{bondi_envisioning_2021}
Bondi, E.; Xu, L.; Acosta-Navas, D.; and Killian, J.~A. 2021.
\newblock Envisioning {Communities}: {A} {Participatory} {Approach} {Towards} {AI} for {Social} {Good}.
\newblock In \emph{Proceedings of the 2021 {AAAI}/{ACM} {Conference} on {AI}, {Ethics}, and {Society}}, {AIES} '21, 425--436. New York, NY, USA: Association for Computing Machinery.
\newblock ISBN 978-1-4503-8473-5.

\bibitem[{Braun and Clarke(2006)}]{braun_using_2006}
Braun, V.; and Clarke, V. 2006.
\newblock Using thematic analysis in psychology.
\newblock \emph{Qualitative research in psychology}, 3(2): 77--101.
\newblock Publisher: Taylor \& Francis.

\bibitem[{Bødker and Kyng(2018)}]{bodker_participatory_2018}
Bødker, S.; and Kyng, M. 2018.
\newblock Participatory design that matters—{Facing} the big issues.
\newblock \emph{ACM Transactions on Computer-Human Interaction (TOCHI)}, 25(1): 1--31.
\newblock Publisher: ACM New York, NY, USA.

\bibitem[{Calacci(2022)}]{calacci_organizing_2022}
Calacci, D. 2022.
\newblock Organizing in the end of employment: information sharing, data stewardship, and digital workerism.
\newblock In \emph{Proceedings of the 1st {Annual} {Meeting} of the {Symposium} on {Human}-{Computer} {Interaction} for {Work}}, 1--9.

\bibitem[{Calacci and Pentland(2022)}]{calacci_bargaining_2022}
Calacci, D.; and Pentland, A. 2022.
\newblock Bargaining with the black-box: {Designing} and deploying worker-centric tools to audit algorithmic management.
\newblock \emph{Proceedings of the ACM on Human-Computer Interaction}, 6(CSCW2): 1--24.
\newblock Publisher: ACM New York, NY, USA.

\bibitem[{Cheng et~al.(2021)Cheng, Stapleton, Wang, Bullock, Chouldechova, Wu, and Zhu}]{cheng_soliciting_2021}
Cheng, H.-F.; Stapleton, L.; Wang, R.; Bullock, P.; Chouldechova, A.; Wu, Z. S.~S.; and Zhu, H. 2021.
\newblock Soliciting stakeholders’ fairness notions in child maltreatment predictive systems.
\newblock In \emph{Proceedings of the 2021 {CHI} {Conference} on {Human} {Factors} in {Computing} {Systems}}, 1--17.

\bibitem[{Cooper and Zafiroglu(2024)}]{cooper_fitting_2024}
Cooper, N.; and Zafiroglu, A. 2024.
\newblock From {Fitting} {Participation} to {Forging} {Relationships}: {The} {Art} of {Participatory} {ML}.
\newblock \emph{arXiv preprint arXiv:2403.06431}.

\bibitem[{Corbett, Denton, and Erete(2023)}]{corbett_power_2023}
Corbett, E.; Denton, E.; and Erete, S. 2023.
\newblock Power and public participation in ai.
\newblock In \emph{Proceedings of the 3rd {ACM} {Conference} on {Equity} and {Access} in {Algorithms}, {Mechanisms}, and {Optimization}}, 1--13.

\bibitem[{Costanza-Chock(2020{\natexlab{a}})}]{costanza-chock_design_2020}
Costanza-Chock, S. 2020{\natexlab{a}}.
\newblock \emph{Design justice: {Community}-led practices to build the worlds we need}.
\newblock The MIT Press.

\bibitem[{Costanza-Chock(2020{\natexlab{b}})}]{costanza-chock_introduction_2020}
Costanza-Chock, S. 2020{\natexlab{b}}.
\newblock Introduction: \#{TravelingWhileTrans}, {Design} {Justice}, and {Escape} from the {Matrix} of {Domination}.
\newblock In \emph{Design {Justice}}. 1 edition.

\bibitem[{De, Kanthawala, and Maddox(2025)}]{de2025gets}
De, A.; Kanthawala, S.; and Maddox, J. 2025.
\newblock Who Gets Heard? Calling Out the" Hard-to-Reach" Myth for Non-WEIRD Populations’ Recruitment and Involvement in Research.
\newblock In \emph{Proceedings of the 2025 ACM Conference on Fairness, Accountability, and Transparency}, 855--867.

\bibitem[{Delgado, Barocas, and Levy(2022)}]{delgado_uncommon_2022}
Delgado, F.; Barocas, S.; and Levy, K. 2022.
\newblock An uncommon task: {Participatory} design in legal {AI}.
\newblock \emph{Proceedings of the ACM on Human-Computer Interaction}, 6(CSCW1): 1--23.
\newblock Publisher: ACM New York, NY, USA.

\bibitem[{Delgado et~al.(2023)Delgado, Yang, Madaio, and Yang}]{delgado_participatory_2023}
Delgado, F.; Yang, S.; Madaio, M.; and Yang, Q. 2023.
\newblock The participatory turn in ai design: {Theoretical} foundations and the current state of practice.
\newblock In \emph{Proceedings of the 3rd {ACM} {Conference} on {Equity} and {Access} in {Algorithms}, {Mechanisms}, and {Optimization}}, 1--23.

\bibitem[{Deng et~al.(2009)Deng, Dong, Socher, Li, Li, and Fei-Fei}]{deng_imagenet_2009}
Deng, J.; Dong, W.; Socher, R.; Li, L.-J.; Li, K.; and Fei-Fei, L. 2009.
\newblock Imagenet: {A} large-scale hierarchical image database.
\newblock In \emph{2009 {IEEE} conference on computer vision and pattern recognition}, 248--255. Ieee.

\bibitem[{Diaz and Smith(2024)}]{diaz2024makes}
Diaz, M.; and Smith, A.~D. 2024.
\newblock What Makes An Expert? Reviewing How ML Researchers Define" Expert".
\newblock In \emph{Proceedings of the AAAI/ACM Conference on AI, Ethics, and Society}, volume~7, 358--370.

\bibitem[{D'ignazio and Klein(2023)}]{dignazio_data_2023}
D'ignazio, C.; and Klein, L.~F. 2023.
\newblock \emph{Data feminism}.
\newblock MIT press.

\bibitem[{Eitzel et~al.(2021)Eitzel, Solera, Hove, Wilson, Ndlovu, Ndlovu, Changarara, Ndlovu, Neves, Chirindira, and {others}}]{eitzel_assessing_2021}
Eitzel, M.; Solera, J.; Hove, E.~M.; Wilson, K.; Ndlovu, A.~M.; Ndlovu, D.; Changarara, A.; Ndlovu, A.; Neves, K.; Chirindira, A.; and {others}. 2021.
\newblock Assessing the potential of participatory modeling for decolonial restoration of an agro-pastoral system in rural {Zimbabwe}.
\newblock \emph{Citizen Science: Theory and Practice}, 6(1).

\bibitem[{Erete et~al.(2025)Erete, Corbett, Smith-Walker, Cunningham, Gatz, Park, Perry, Wilcox, and Denton}]{erete_towards_2025}
Erete, S.; Corbett, E.; Smith-Walker, N.; Cunningham, J.~L.; Gatz, E.; Park, T.; Perry, T.; Wilcox, L.; and Denton, R. 2025.
\newblock Towards {Equitable} {Community}-{Industry} {Collaborations}: {Understanding} the {Experiences} of {Nonprofits}' {Collaborations} with {Tech} {Companies}.
\newblock \emph{Proceedings of the ACM on Human-Computer Interaction}, 9(2): 1--31.
\newblock Publisher: ACM New York, NY, USA.

\bibitem[{Erete, Rankin, and Thomas(2021)}]{erete_i_2021}
Erete, S.; Rankin, Y.~A.; and Thomas, J.~O. 2021.
\newblock I can't breathe: {Reflections} from {Black} women in {CSCW} and {HCI}.
\newblock \emph{Proceedings of the ACM on Human-Computer Interaction}, 4(CSCW3): 1--23.
\newblock Publisher: ACM New York, NY, USA.

\bibitem[{Feffer et~al.(2023)Feffer, Skirpan, Lipton, and Heidari}]{feffer_preference_2023}
Feffer, M.; Skirpan, M.; Lipton, Z.; and Heidari, H. 2023.
\newblock From {Preference} {Elicitation} to {Participatory} {ML}: {A} {Critical} {Survey} \& {Guidelines} for {Future} {Research}.
\newblock In \emph{Proceedings of the 2023 {AAAI}/{ACM} {Conference} on {AI}, {Ethics}, and {Society}}, 38--48. Montr{\textbackslash}'\{e\}al QC Canada: ACM.
\newblock ISBN 9798400702310.

\bibitem[{Finn et~al.(2022)Finn, Jones, Mahelona, Duncan, and Leoni}]{finn_developing_2022}
Finn, A.; Jones, P.-L.; Mahelona, K.; Duncan, S.; and Leoni, G. 2022.
\newblock Developing a {Part}-{Of}-{Speech} tagger for te reo {Māori}.
\newblock In \emph{Proceedings of the {Fifth} {Workshop} on the {Use} of {Computational} {Methods} in the {Study} of {Endangered} {Languages}}, 93--98.

\bibitem[{Flanigan et~al.(2021)Flanigan, Gölz, Gupta, Hennig, and Procaccia}]{flanigan_fair_2021}
Flanigan, B.; Gölz, P.; Gupta, A.; Hennig, B.; and Procaccia, A.~D. 2021.
\newblock Fair algorithms for selecting citizens’ assemblies.
\newblock \emph{Nature}, 596(7873): 548--552.

\bibitem[{Frauenberger et~al.(2015)Frauenberger, Good, Fitzpatrick, and Iversen}]{frauenberger_pursuit_2015}
Frauenberger, C.; Good, J.; Fitzpatrick, G.; and Iversen, O.~S. 2015.
\newblock In pursuit of rigour and accountability in participatory design.
\newblock \emph{International journal of human-computer studies}, 74: 93--106.
\newblock Publisher: Elsevier.

\bibitem[{Freedman et~al.(2020)Freedman, Borg, Sinnott-Armstrong, Dickerson, and Conitzer}]{freedman_adapting_2020}
Freedman, R.; Borg, J.~S.; Sinnott-Armstrong, W.; Dickerson, J.~P.; and Conitzer, V. 2020.
\newblock Adapting a kidney exchange algorithm to align with human values.
\newblock \emph{Artificial Intelligence}, 283: 103261.
\newblock Publisher: Elsevier.

\bibitem[{Gibson and Abrams(2003)}]{gibson_racial_2003}
Gibson, P.; and Abrams, L. 2003.
\newblock Racial difference in engaging, recruiting, and interviewing {African} {American} women in qualitative research.
\newblock \emph{Qualitative Social Work}, 2(4): 457--476.
\newblock Publisher: Sage Publications.

\bibitem[{Gilman(2023)}]{gilman_democratizing_2023}
Gilman, M.~E. 2023.
\newblock Democratizing {AI}: {Principles} for {Meaningful} {Public} {Participation}.
\newblock \emph{Data \& Society}.

\bibitem[{Groves et~al.(2023)Groves, Peppin, Strait, and Brennan}]{groves_going_2023}
Groves, L.; Peppin, A.; Strait, A.; and Brennan, J. 2023.
\newblock Going public: the role of public participation approaches in commercial {AI} labs.
\newblock In \emph{Proceedings of the 2023 {ACM} {Conference} on {Fairness}, {Accountability}, and {Transparency}}, 1162--1173.

\bibitem[{Halfaker and Geiger(2020)}]{halfaker_ores_2020}
Halfaker, A.; and Geiger, R.~S. 2020.
\newblock Ores: {Lowering} barriers with participatory machine learning in wikipedia.
\newblock \emph{Proceedings of the ACM on Human-Computer Interaction}, 4(CSCW2): 1--37.
\newblock Publisher: ACM New York, NY, USA.

\bibitem[{Harrington, Erete, and Piper(2019)}]{harrington_deconstructing_2019}
Harrington, C.; Erete, S.; and Piper, A.~M. 2019.
\newblock Deconstructing community-based collaborative design: {Towards} more equitable participatory design engagements.
\newblock \emph{Proceedings of the ACM on human-computer interaction}, 3(CSCW): 1--25.
\newblock Publisher: ACM New York, NY, USA.

\bibitem[{Heerman et~al.(2017)Heerman, Jackson, Roumie, Harris, Rosenbloom, Pulley, Wilkins, Williams, Crenshaw, Leak, and {others}}]{heerman_recruitment_2017}
Heerman, W.~J.; Jackson, N.; Roumie, C.~L.; Harris, P.~A.; Rosenbloom, S.~T.; Pulley, J.; Wilkins, C.~H.; Williams, N.~A.; Crenshaw, D.; Leak, C.; and {others}. 2017.
\newblock Recruitment methods for survey research: findings from the mid-south clinical data research network.
\newblock \emph{Contemporary clinical trials}, 62: 50--55.
\newblock Publisher: Elsevier.

\bibitem[{Hollingshead and Poole(2012)}]{hollingshead_research_2012}
Hollingshead, A.; and Poole, M.~S. 2012.
\newblock \emph{Research methods for studying groups and teams: {A} guide to approaches, tools, and technologies}.
\newblock Routledge.

\bibitem[{Imran et~al.(2020)Imran, Alam, Qazi, Peterson, and Ofli}]{imran_rapid_2020}
Imran, M.; Alam, F.; Qazi, U.; Peterson, S.; and Ofli, F. 2020.
\newblock Rapid damage assessment using social media images by combining human and machine intelligence.
\newblock \emph{arXiv preprint arXiv:2004.06675}.

\bibitem[{Irani et~al.(2010)Irani, Vertesi, Dourish, Philip, and Grinter}]{irani_postcolonial_2010}
Irani, L.; Vertesi, J.; Dourish, P.; Philip, K.; and Grinter, R.~E. 2010.
\newblock Postcolonial computing: a lens on design and development.
\newblock In \emph{Proceedings of the {SIGCHI} conference on human factors in computing systems}, 1311--1320.

\bibitem[{Irani and Silberman(2013)}]{irani_turkopticon_2013}
Irani, L.~C.; and Silberman, M.~S. 2013.
\newblock Turkopticon: {Interrupting} worker invisibility in amazon mechanical turk.
\newblock In \emph{Proceedings of the {SIGCHI} conference on human factors in computing systems}, 611--620.

\bibitem[{Ismail et~al.(2023)Ismail, Thakkar, Madhiwalla, and Kumar}]{ismail_public_2023}
Ismail, A.; Thakkar, D.; Madhiwalla, N.; and Kumar, N. 2023.
\newblock Public health calls for/with {AI}: an ethnographic perspective.
\newblock \emph{Proceedings of the ACM on Human-Computer Interaction}, 7(CSCW2): 1--26.
\newblock Publisher: ACM New York, NY, USA.

\bibitem[{Kallina, Bohn{\'e}, and Singh(2025)}]{kallina2025stakeholder}
Kallina, E.; Bohn{\'e}, T.; and Singh, J. 2025.
\newblock Stakeholder Participation for Responsible AI Development: Disconnects Between Guidance and Current Practice.
\newblock In \emph{Proceedings of the 2025 ACM Conference on Fairness, Accountability, and Transparency}, 1060--1079.

\bibitem[{Kallina and Singh(2024)}]{kallina_stakeholder_2024}
Kallina, E.; and Singh, J. 2024.
\newblock Stakeholder {Involvement} for {Responsible} {AI} {Development}: {A} {Process} {Framework}.
\newblock In \emph{Proceedings of the 4th {ACM} {Conference} on {Equity} and {Access} in {Algorithms}, {Mechanisms}, and {Optimization}}, 1--14. San Luis Potosi Mexico: ACM.
\newblock ISBN 9798400712227.

\bibitem[{Kamineni et~al.(2023)Kamineni, Ötleş, Oh, Rao, Young, Li, West, Hooper, Shenoy, Guttag, and {others}}]{kamineni_prospective_2023}
Kamineni, M.; Ötleş, E.; Oh, J.; Rao, K.; Young, V.~B.; Li, B.~Y.; West, L.~R.; Hooper, D.~C.; Shenoy, E.~S.; Guttag, J.~G.; and {others}. 2023.
\newblock Prospective evaluation of data-driven models to predict daily risk of {Clostridioides} difficile infection at 2 large academic health centers.
\newblock \emph{Infection Control \& Hospital Epidemiology}, 44(7): 1163--1166.
\newblock Publisher: Cambridge University Press.

\bibitem[{Katell et~al.(2020)Katell, Young, Dailey, Herman, Guetler, Tam, Bintz, Raz, and Krafft}]{katell_toward_2020}
Katell, M.; Young, M.; Dailey, D.; Herman, B.; Guetler, V.; Tam, A.; Bintz, C.; Raz, D.; and Krafft, P. 2020.
\newblock Toward situated interventions for algorithmic equity: lessons from the field.
\newblock In \emph{Proceedings of the 2020 conference on fairness, accountability, and transparency}, 45--55.

\bibitem[{Kelling et~al.(2012)Kelling, Gerbracht, Fink, Lagoze, Wong, Yu, Damoulas, and Gomes}]{kelling_ebird_2012}
Kelling, S.; Gerbracht, J.; Fink, D.; Lagoze, C.; Wong, W.-K.; Yu, J.~Y.; Damoulas, T.; and Gomes, C. 2012.
\newblock ebird: {A} human/computer learning network for biodiversity conservation and research.
\newblock In \emph{Proceedings of the {AAAI} {Conference} on {Artificial} {Intelligence}}, volume~26, 2229--2236.
\newblock Issue: 2.

\bibitem[{Klein and D'Ignazio(2024)}]{klein_data_2024}
Klein, L.; and D'Ignazio, C. 2024.
\newblock Data {Feminism} for {AI}.
\newblock In \emph{Proceedings of the 2024 {ACM} {Conference} on {Fairness}, {Accountability}, and {Transparency}}, {FAccT} '24, 100--112. New York, NY, USA: Association for Computing Machinery.
\newblock ISBN 9798400704505.

\bibitem[{Kulynych et~al.(2020)Kulynych, Madras, Milli, Raji, Zhou, and Zemel}]{kulynych_participatory_2020}
Kulynych, B.; Madras, D.; Milli, S.; Raji, I.~D.; Zhou, A.; and Zemel, R. 2020.
\newblock Participatory {Approaches} to {Machine} {Learning}.
\newblock Published: International Conference on Machine Learning Workshop.

\bibitem[{Kuo et~al.(2023)Kuo, Shen, Geum, Jones, Hong, Zhu, and Holstein}]{kuo_understanding_2023}
Kuo, T.-S.; Shen, H.; Geum, J.; Jones, N.; Hong, J.~I.; Zhu, H.; and Holstein, K. 2023.
\newblock Understanding frontline workers’ and unhoused individuals’ perspectives on ai used in homeless services.
\newblock In \emph{Proceedings of the 2023 {CHI} {Conference} on {Human} {Factors} in {Computing} {Systems}}, 1--17.

\bibitem[{Langer et~al.(2021)Langer, Castro, Chen, Davis, Joseph, Kim, Larkey, Lee, Petrov, Reifsnider, Youngstedt, and Shaibi}]{langer_recruitment_2021}
Langer, S.~L.; Castro, F.~G.; Chen, A. C.-C.; Davis, K.~C.; Joseph, R.~P.; Kim, W.~S.; Larkey, L.; Lee, R.~E.; Petrov, M.~E.; Reifsnider, E.; Youngstedt, S.~D.; and Shaibi, G.~Q. 2021.
\newblock Recruitment and {Retention} of {Underrepresented} and {Vulnerable} {Populations} to {Research}.
\newblock \emph{Public health nursing (Boston, Mass.)}, 38(6): 1102--1115.

\bibitem[{Le~Dantec and Fox(2015)}]{le_dantec_strangers_2015}
Le~Dantec, C.~A.; and Fox, S. 2015.
\newblock Strangers at the gate: {Gaining} access, building rapport, and co-constructing community-based research.
\newblock In \emph{Proceedings of the 18th {ACM} conference on computer supported cooperative work \& social computing}, 1348--1358.

\bibitem[{Lee et~al.(2019)Lee, Kusbit, Kahng, Kim, Yuan, Chan, See, Noothigattu, Lee, Psomas, and {others}}]{lee_webuildai_2019}
Lee, M.~K.; Kusbit, D.; Kahng, A.; Kim, J.~T.; Yuan, X.; Chan, A.; See, D.; Noothigattu, R.; Lee, S.; Psomas, A.; and {others}. 2019.
\newblock {WeBuildAI}: {Participatory} framework for algorithmic governance.
\newblock \emph{Proceedings of the ACM on human-computer interaction}, 3(CSCW): 1--35.
\newblock Publisher: ACM New York, NY, USA.

\bibitem[{Leslie, Khayatzadeh-Mahani, and MacKean(2019)}]{leslie_recruitment_2019}
Leslie, M.; Khayatzadeh-Mahani, A.; and MacKean, G. 2019.
\newblock Recruitment of caregivers into health services research: lessons from a user-centred design study.
\newblock \emph{Research involvement and engagement}, 5: 1--9.
\newblock Publisher: Springer.

\bibitem[{Li et~al.(2023)Li, Kingsley, Fan, Sinha, Wai, Lee, Shen, Eslami, and Hong}]{li_participation_2023}
Li, R.; Kingsley, S.; Fan, C.; Sinha, P.; Wai, N.; Lee, J.; Shen, H.; Eslami, M.; and Hong, J. 2023.
\newblock Participation and {Division} of {Labor} in {User}-{Driven} {Algorithm} {Audits}: {How} {Do} {Everyday} {Users} {Work} together to {Surface} {Algorithmic} {Harms}?
\newblock In \emph{Proceedings of the 2023 {CHI} {Conference} on {Human} {Factors} in {Computing} {Systems}}, 1--19.

\bibitem[{Liao and Muller(2019)}]{liao_enabling_2019}
Liao, Q.~V.; and Muller, M. 2019.
\newblock Enabling value sensitive {AI} systems through participatory design fictions.
\newblock \emph{arXiv preprint arXiv:1912.07381}.

\bibitem[{Lin et~al.(2024)Lin, Karusala, Okolo, D'Ignazio, and Gajos}]{lin_come_2024}
Lin, H.; Karusala, N.; Okolo, C.~T.; D'Ignazio, C.; and Gajos, K.~Z. 2024.
\newblock "{Come} to us first": {Centering} {Community} {Organizations} in {Artificial} {Intelligence} for {Social} {Good} {Partnerships}.
\newblock ArXiv:2409.06814 [cs].

\bibitem[{Mancera et~al.(2022)Mancera, Sy, Williams, and Hargreaves}]{mancera_utilizing_2022}
Mancera, B.~M.; Sy, A.; Williams, C.~D.; and Hargreaves, M.~K. 2022.
\newblock Utilizing a social-ecological health promotion framework to engage diverse populations for recruitment in the {All} of {Us} research program.
\newblock \emph{Journal of Community Engagement and Scholarship}, 13(2).

\bibitem[{Marian(2023)}]{marian_algorithmic_2023}
Marian, A. 2023.
\newblock Algorithmic transparency and accountability through crowdsourcing: {A} study of the nyc school admission lottery.
\newblock In \emph{Proceedings of the 2023 {ACM} {Conference} on {Fairness}, {Accountability}, and {Transparency}}, 434--443.

\bibitem[{Miceli et~al.(2021)Miceli, Yang, Naudts, Schuessler, Serbanescu, and Hanna}]{miceli_documenting_2021}
Miceli, M.; Yang, T.; Naudts, L.; Schuessler, M.; Serbanescu, D.; and Hanna, A. 2021.
\newblock Documenting computer vision datasets: an invitation to reflexive data practices.
\newblock In \emph{Proceedings of the 2021 {ACM} {Conference} on {Fairness}, {Accountability}, and {Transparency}}, 161--172.

\bibitem[{Mohamed, Png, and Isaac(2020)}]{mohamed_decolonial_2020}
Mohamed, S.; Png, M.-T.; and Isaac, W. 2020.
\newblock Decolonial {AI}: {Decolonial} theory as sociotechnical foresight in artificial intelligence.
\newblock \emph{Philosophy \& Technology}, 33: 659--684.
\newblock Publisher: Springer.

\bibitem[{Nabatchi et~al.(2012)Nabatchi, Gastil, Weiksner, and Leighninger}]{nabatchi_democracy_2012}
Nabatchi, T.; Gastil, J.; Weiksner, G.~M.; and Leighninger, M. 2012.
\newblock \emph{Democracy in motion: {Evaluating} the practice and impact of deliberative civic engagement}.
\newblock Oxford University Press.

\bibitem[{Negrin et~al.(2022)Negrin, Slaughter, Dahlke, and Olson}]{negrin_successful_2022}
Negrin, K.~A.; Slaughter, S.~E.; Dahlke, S.; and Olson, J. 2022.
\newblock Successful recruitment to qualitative research: {A} critical reflection.
\newblock \emph{International Journal of Qualitative Methods}, 21: 16094069221119576.
\newblock Publisher: SAGE Publications Sage CA: Los Angeles, CA.

\bibitem[{Nekoto et~al.(2020)Nekoto, Marivate, Matsila, Fasubaa, Kolawole, Fagbohungbe, Akinola, Muhammad, Kabongo, Osei, and {others}}]{nekoto_participatory_2020}
Nekoto, W.; Marivate, V.; Matsila, T.; Fasubaa, T.; Kolawole, T.; Fagbohungbe, T.; Akinola, S.~O.; Muhammad, S.~H.; Kabongo, S.; Osei, S.; and {others}. 2020.
\newblock Participatory research for low-resourced machine translation: {A} case study in african languages.
\newblock \emph{arXiv preprint arXiv:2010.02353}.

\bibitem[{Newman et~al.(2022)Newman, Immorlica, Leyton-Brown, Lucier, Quinn, and Ssekibuule}]{newman_kudu_nodate}
Newman, N.; Immorlica, N.; Leyton-Brown, K.; Lucier, B.; Quinn, J.; and Ssekibuule, R. 2022.
\newblock Kudu: {An} {Electronic} {Agricultural} {Marketplace} in {Uganda}.

\bibitem[{Obermeyer et~al.(2019)Obermeyer, Powers, Vogeli, and Mullainathan}]{obermeyer_dissecting_2019}
Obermeyer, Z.; Powers, B.; Vogeli, C.; and Mullainathan, S. 2019.
\newblock Dissecting racial bias in an algorithm used to manage the health of populations.
\newblock \emph{Science}, 366(6464): 447--453.
\newblock Publisher: American Association for the Advancement of Science.

\bibitem[{O’Brien et~al.(2022)O’Brien, Brewer, Jones, Corkhum, and Rizo}]{obrien_rigor_2022}
O’Brien, J.~E.; Brewer, K.~B.; Jones, L.~M.; Corkhum, J.; and Rizo, C.~F. 2022.
\newblock Rigor and respect: recruitment strategies for engaging vulnerable populations in research.
\newblock \emph{Journal of interpersonal violence}, 37(17-18): NP17052--NP17072.
\newblock Publisher: SAGE Publications Sage CA: Los Angeles, CA.

\bibitem[{Perrault et~al.(2020)Perrault, Fang, Sinha, and Tambe}]{perrault_artificial_2020}
Perrault, A.; Fang, F.; Sinha, A.; and Tambe, M. 2020.
\newblock Artificial intelligence for social impact: {Learning} and planning in the data-to-deployment pipeline.
\newblock \emph{AI Magazine}, 41(4): 3--16.

\bibitem[{Puussaar et~al.(2022)Puussaar, Montague, Peacock, Nappey, Anderson, Jonczyk, Wright, and James}]{puussaar_sensemystreet_2022}
Puussaar, A.; Montague, K.; Peacock, S.; Nappey, T.; Anderson, R.; Jonczyk, J.; Wright, P.; and James, P. 2022.
\newblock {SenseMyStreet}: {Sensor} {Commissioning} {Toolkit} for {Communities}.
\newblock \emph{Proceedings of the ACM on Human-Computer Interaction}, 6(CSCW2): 1--26.
\newblock Publisher: ACM New York, NY, USA.

\bibitem[{Queerinai et~al.(2023)Queerinai, Ovalle, Subramonian, Singh, Voelcker, Sutherland, Locatelli, Breznik, Klubicka, Yuan, J, Zhang, Shriram, Lehman, Soldaini, Sap, Deisenroth, Pacheco, Ryskina, Mundt, Agarwal, Mclean, Xu, Pranav, Korpan, Ray, Mathew, Arora, John, Anand, Agrawal, Agnew, Long, Wang, Talat, Ghosh, Dennler, Noseworthy, Jha, Baylor, Joshi, Bilenko, Mcnamara, Gontijo-Lopes, Markham, Dong, Kay, Saraswat, Vytla, and Stark}]{queerinai_queer_2023}
Queerinai, O.~O.; Ovalle, A.; Subramonian, A.; Singh, A.; Voelcker, C.; Sutherland, D.~J.; Locatelli, D.; Breznik, E.; Klubicka, F.; Yuan, H.; J, H.; Zhang, H.; Shriram, J.; Lehman, K.; Soldaini, L.; Sap, M.; Deisenroth, M.~P.; Pacheco, M.~L.; Ryskina, M.; Mundt, M.; Agarwal, M.; Mclean, N.; Xu, P.; Pranav, A.; Korpan, R.; Ray, R.; Mathew, S.; Arora, S.; John, S.; Anand, T.; Agrawal, V.; Agnew, W.; Long, Y.; Wang, Z.~J.; Talat, Z.; Ghosh, A.; Dennler, N.; Noseworthy, M.; Jha, S.; Baylor, E.; Joshi, A.; Bilenko, N.~Y.; Mcnamara, A.; Gontijo-Lopes, R.; Markham, A.; Dong, E.; Kay, J.; Saraswat, M.; Vytla, N.; and Stark, L. 2023.
\newblock Queer {In} {AI}: {A} {Case} {Study} in {Community}-{Led} {Participatory} {AI}.
\newblock In \emph{2023 {ACM} {Conference} on {Fairness}, {Accountability}, and {Transparency}}, 1882--1895. Chicago IL USA: ACM.
\newblock ISBN 9798400701924.

\bibitem[{Roscoe(2021)}]{roscoe_designing_2021}
Roscoe, R. 2021.
\newblock Designing for diversity: {Inclusive} {Sampling}.
\newblock \emph{Ergodesign \& HCI}, 9.

\bibitem[{Sendak et~al.(2020)Sendak, Elish, Gao, Futoma, Ratliff, Nichols, Bedoya, Balu, and O'Brien}]{sendak__2020}
Sendak, M.; Elish, M.~C.; Gao, M.; Futoma, J.; Ratliff, W.; Nichols, M.; Bedoya, A.; Balu, S.; and O'Brien, C. 2020.
\newblock " {The} human body is a black box" supporting clinical decision-making with deep learning.
\newblock In \emph{Proceedings of the 2020 conference on fairness, accountability, and transparency}, 99--109.

\bibitem[{Sharp and Anderson(2010)}]{sharp_online_2010}
Sharp, A.; and Anderson, K. 2010.
\newblock Online citizen panels as an advance in research and consultation: {A} {Review} of pilot results.
\newblock \emph{Commonwealth Journal of Local Governance}, (6): 33--54.

\bibitem[{Shen et~al.(2022)Shen, Wang, Deng, Brusse, Velgersdijk, and Zhu}]{shen_model_2022}
Shen, H.; Wang, L.; Deng, W.~H.; Brusse, C.; Velgersdijk, R.; and Zhu, H. 2022.
\newblock The model card authoring toolkit: {Toward} community-centered, deliberation-driven {AI} design.
\newblock In \emph{Proceedings of the 2022 {ACM} {Conference} on {Fairness}, {Accountability}, and {Transparency}}, 440--451.

\bibitem[{Shi et~al.(2020)Shi, Yuan, Lo, Lizarondo, and Fang}]{shi_improving_2020}
Shi, Z.~R.; Yuan, Y.; Lo, K.; Lizarondo, L.; and Fang, F. 2020.
\newblock Improving efficiency of volunteer-based food rescue operations.
\newblock In \emph{Proceedings of the {AAAI} {Conference} on {Artificial} {Intelligence}}, volume~34, 13369--13375.
\newblock Issue: 08.

\bibitem[{Sinders(2020)}]{sinders_feminist_2020}
Sinders, C. 2020.
\newblock Feminist data set.
\newblock \emph{Clinic for Open Source Arts. https://carolinesinders. com/wp-con tent/uploads/2020/05/Feminist-Data-Set-Final-Draft-2020-0517. pd f}.

\bibitem[{Sloane et~al.(2022)Sloane, Moss, Awomolo, and Forlano}]{sloane_participation_2022}
Sloane, M.; Moss, E.; Awomolo, O.; and Forlano, L. 2022.
\newblock Participation is not a design fix for machine learning.
\newblock In \emph{Proceedings of the 2nd {ACM} {Conference} on {Equity} and {Access} in {Algorithms}, {Mechanisms}, and {Optimization}}, 1--6.

\bibitem[{Small and Calarco(2022)}]{small_qualitative_2022}
Small, M.~L.; and Calarco, J.~M. 2022.
\newblock \emph{Qualitative literacy: {A} guide to evaluating ethnographic and interview research}.
\newblock Univ of California Press.

\bibitem[{Spinuzzi(2005)}]{spinuzzi_methodology_nodate}
Spinuzzi, C. 2005.
\newblock The Methodology of Participatory Design.

\bibitem[{Sum et~al.(2025)Sum, Zhi, Cook, Cooper, Lozano, Johnson, Perez, Ghani, Skirpan, Eslami, and {others}}]{sum__2025}
Sum, C.~M.; Zhi, J.; Cook, A.~N.; Cooper, P.~J.; Lozano, A.; Johnson, T.; Perez, J.; Ghani, R.; Skirpan, M.; Eslami, M.; and {others}. 2025.
\newblock " {You}'re in a {Ferrari}. {I}'m {Waiting} for the {Bus}": {Confronting} {Tensions} in {Community}-{University} {Partnerships}.
\newblock \emph{Proceedings of the ACM on Human-Computer Interaction}, 9(2): 1--28.
\newblock Publisher: ACM New York, NY, USA.

\bibitem[{Suresh et~al.(2022)Suresh, Movva, Dogan, Bhargava, Crux{\^e}n, Cuba, Taurino, So, and D'Ignazio}]{suresh_towards_2022}
Suresh, H.; Movva, R.; Dogan, A.~L.; Bhargava, R.; Crux{\^e}n, I.; Cuba, {\'{A}}.~M.; Taurino, G.; So, W.; and D'Ignazio, C. 2022.
\newblock Towards intersectional feminist and participatory {ML}: {A} case study in supporting {Feminicide} {Counterdata} {Collection}.
\newblock In \emph{Proceedings of the 2022 {ACM} {Conference} on {Fairness}, {Accountability}, and {Transparency}}, 667--678.

\bibitem[{Swanson et~al.(2015)Swanson, Kosmala, Lintott, Simpson, Smith, and Packer}]{swanson_snapshot_2015}
Swanson, A.; Kosmala, M.; Lintott, C.; Simpson, R.; Smith, A.; and Packer, C. 2015.
\newblock Snapshot {Serengeti}, high-frequency annotated camera trap images of 40 mammalian species in an {African} savanna.
\newblock \emph{Scientific Data}, 2(1): 150026.
\newblock Publisher: Nature Publishing Group.

\bibitem[{Tambe, Fang, and Wilder(2022)}]{tambe_ai_2022}
Tambe, M.; Fang, F.; and Wilder, B., eds. 2022.
\newblock \emph{{AI} for {Social} {Impact}}.

\bibitem[{Tandon et~al.(2022)Tandon, Khovanskaya, Arcilla, Hussein, Zschiesche, and Irani}]{tandon_hostile_2022}
Tandon, U.; Khovanskaya, V.; Arcilla, E.; Hussein, M.~H.; Zschiesche, P.; and Irani, L. 2022.
\newblock Hostile ecologies: {Navigating} the barriers to community-led innovation.
\newblock \emph{Proceedings of the ACM on Human-Computer Interaction}, 6(CSCW2): 1--26.
\newblock Publisher: ACM New York, NY, USA.

\bibitem[{Tuia et~al.(2022)Tuia, Kellenberger, Beery, Costelloe, Zuffi, Risse, Mathis, Mathis, Van~Langevelde, Burghardt, and {others}}]{tuia_perspectives_2022}
Tuia, D.; Kellenberger, B.; Beery, S.; Costelloe, B.~R.; Zuffi, S.; Risse, B.; Mathis, A.; Mathis, M.~W.; Van~Langevelde, F.; Burghardt, T.; and {others}. 2022.
\newblock Perspectives in machine learning for wildlife conservation.
\newblock \emph{Nature communications}, 13(1): 792.
\newblock Publisher: Nature Publishing Group UK London.

\bibitem[{{UNHCR Innovation Service}(2019)}]{service_goat_2019}
{UNHCR Innovation Service}. 2019.
\newblock A Goat Story.
\newblock \url{https://medium.com/unhcr-innovation-service/a-goat-story-3ed6bdd2b237}.

\bibitem[{{University of Michigan STPP program and Detroit Disability Power and the Detroit Justice Center;}(2024)}]{university_of_michigan_stpp_program_and_detroit_disability_power_and_the_detroit_justice_center_community_2024}
{University of Michigan STPP program and Detroit Disability Power and the Detroit Justice Center;}. 2024.
\newblock Community {Partnerships} {Playbook}.

\bibitem[{Vines et~al.(2013)Vines, Clarke, Wright, McCarthy, and Olivier}]{vines_configuring_2013}
Vines, J.; Clarke, R.; Wright, P.; McCarthy, J.; and Olivier, P. 2013.
\newblock Configuring participation: on how we involve people in design.
\newblock In \emph{Proceedings of the {SIGCHI} {Conference} on {Human} {Factors} in {Computing} {Systems}}, 429--438. Paris France: ACM.
\newblock ISBN 978-1-4503-1899-0.

\bibitem[{Wilder et~al.(2021)Wilder, Onasch-Vera, Diguiseppi, Petering, Hill, Yadav, Rice, and Tambe}]{wilder_clinical_2021}
Wilder, B.; Onasch-Vera, L.; Diguiseppi, G.; Petering, R.; Hill, C.; Yadav, A.; Rice, E.; and Tambe, M. 2021.
\newblock Clinical trial of an {AI}-augmented intervention for {HIV} prevention in youth experiencing homelessness.
\newblock In \emph{Proceedings of the {AAAI} {Conference} on {Artificial} {Intelligence}}, volume~35, 14948--14956.
\newblock Issue: 17.

\bibitem[{Wood et~al.(2011)Wood, Sullivan, Iliff, Fink, and Kelling}]{wood_ebird_2011}
Wood, C.; Sullivan, B.; Iliff, M.; Fink, D.; and Kelling, S. 2011.
\newblock {eBird}: engaging birders in science and conservation.
\newblock \emph{PLoS biology}, 9(12): e1001220.
\newblock Publisher: Public Library of Science San Francisco, USA.

\bibitem[{Young, Magassa, and Friedman(2019)}]{young_toward_2019}
Young, M.; Magassa, L.; and Friedman, B. 2019.
\newblock Toward inclusive tech policy design: a method for underrepresented voices to strengthen tech policy documents.
\newblock \emph{Ethics and Information Technology}, 21: 89--103.
\newblock Publisher: Springer.

\bibitem[{Yu et~al.(2020)Yu, Yuan, Terveen, Wu, Forlizzi, and Zhu}]{yu_keeping_2020}
Yu, B.; Yuan, Y.; Terveen, L.; Wu, Z.~S.; Forlizzi, J.; and Zhu, H. 2020.
\newblock Keeping {Designers} in the {Loop}: {Communicating} {Inherent} {Algorithmic} {Trade}-offs {Across} {Multiple} {Objectives}.
\newblock In \emph{Proceedings of the 2020 {ACM} {Designing} {Interactive} {Systems} {Conference}}, 1245--1257. Eindhoven Netherlands: ACM.
\newblock ISBN 978-1-4503-6974-9.

\bibitem[{Zhu et~al.(2018)Zhu, Yu, Halfaker, and Terveen}]{zhu_value-sensitive_2018}
Zhu, H.; Yu, B.; Halfaker, A.; and Terveen, L. 2018.
\newblock Value-{Sensitive} {Algorithm} {Design}: {Method}, {Case} {Study}, and {Lessons}.
\newblock \emph{Proceedings of the ACM on Human-Computer Interaction}, 2(CSCW): 1--23.

\end{thebibliography}

\clearpage
\appendix

\section{Interview Questions} \label{sec:interview_questions}
Prior to the study, we emailed the participants with the specific questions from ParticipAIte (the preliminary database we have developed, Appendix \ref{sec:participaite}) that we planned on asking during the session, as well as the consent form. The interview was a semi-structured interview, which entailed follow-up questions on a case-by-case basis. See the questions outlined below.
\begin{itemize}
\item Introduce the project:
“Hi, my name is [redacted]. Thanks for being here. First, I want to check if you’ve read the consent form, and I can answer any questions you might have about what we’re doing today. I also want to double check that you are consenting to the study protocol we have planned today, including recording.”
“Is it OK for me to hit the record button now?”
“For context, we’re working on a project to develop methodology for Participatory AI. Participatory AI aims to develop AI with interested and impacted communities, and ultimately center their needs and values. We appreciated your work in this space, and would love to learn more. In particular, we are working on a project to help AI practitioners standardize the documentation of participatory processes, like recruitment, so we can distill best practices. I will ask some questions about your work on [specific project], and we’ll discuss how it’s presented in our tool called ParticipAIte. Do you think you could talk about this project?” [Introduce the project]
\item
Next, the participant will view an entry in ParticipAIte that is filled out with details of their project, which will be completed by the authors according to publicly available scientific literature. See attached image for what an entry in ParticipAIte will look like. “This is how your project is presented in our database right now. From your paper, we were able to fill out these details. However, we still want to learn a few more details about the project.” We will ask the relevant questions from the ParticipAIte database that we want more details on. 
Researchers will take note of issues in understanding the content of the database during this step.
After completing the documentation, we will ask some questions about their experience.
\item
“Do you think this kind of information in ParticipAIte is helpful for future research (both your research or others’ research) in this area? Why or why not?”
\item
“Why did you choose to reach out to potential participants for your project?”
\item
“Why did you choose to reach out to participants in this way?”
\item
“Were there any unexpected barriers to doing so, or challenges during the process?”
\item
“What information did you use to develop your recruitment strategy?”
\item
“What did your participants contribute to the project?”
\item
“How did your participants and involved communities benefit from the project?”
\item
“Did you feel that all relevant stakeholders were included, particularly those from traditionally marginalized groups? Why or why not?”
\item
Next, we look through the database together, focusing on one existing case study that is recorded from the AI literature. “Next, I have a different case study I want to show you. Let’s take a look at this together.”
\item
“What are the differences you notice between this case study and yours?”
\item
“Can you tell me more about the difference in recruitment strategies that you notice?”
\item
“Is there anything you would have done differently in your project that we just discussed, now that you’ve seen the case study/entry?”
\item
Finally, we ask for some broad feedback on the tool: “Do you have any suggestions about how we should design the database to help AI practitioners develop strategies for a truly participatory AI?”
\end{itemize}

\section{ParticipAIte, A Database for Community Documentation of Recruitment Strategies}\label{sec:participaite}

One of our motivations and beliefs in writing this paper is that specific documentation can encourage transparency and reproducibility for future projects. As a first step, we created a preliminary database called ParticipAIte, that we hope could inspire further transparency and reproducibility in the future. 

When adding a new case study to ParticipAIte, a form is presented to a user, who we will refer to as a scribe. First, the scribe is asked to identify their role in the project, and list the distinct stakeholder groups involved in the project. Then, for each of the stakeholders, scribes will specify what kind of roles the stakeholder played in the project, who initiated the project, the recruitment strategy that was used to recruit these stakeholders, and any important lessons that were learned from this process of recruitment. 

We now describe each section of the form in detail.

\begin{itemize}
\item \textbf{List the types of stakeholders involved in the project and their role(s)}
These roles are synthesized from codes from our corpus identified in Section 3: \textbf{Clients}, \textbf{Collaborators}, \textbf{Decision Users}, \textbf{Data Creators}, \textbf{Decision Subjects}, and \textbf{Regulators and Governors}.
\end{itemize}

The roles are included to help those filling out the database envision the different groups of people that may have participated in the development of the AI system, whether they were construed as research subjects and formal participants or not. As noted in our coding of ``Who were the different stakeholders involved in the project?", we note that stakeholders can fall under multiple categories, and researchers can engage with more than one stakeholder groups and use different recruitment strategies with each group. This is not necessarily an exhaustive list of stakeholders.

\begin{itemize}
\item \textbf{For each stakeholder, who initiated the recruitment process?} We recall that recruitment can be initiated by researchers or participants, and leave space in this form for indicating who initiated the process of recruitment. 
\end{itemize}
The next question is: 
\begin{itemize}
\item \textbf{What mechanisms did you use to recruit your stakeholder?} The scribe can select an answer from the following four choices, which will lead to additional subquestions that are included for specificity.
\end{itemize}

\begin{enumerate}
\item \textbf{We contacted and worked with an NGO, university, advocacy group, union, etc. to recruit these stakeholders}. This corresponds with the ``Organizational" code. 

\begin{itemize}
\item \textbf{If appropriate to disclose: Which Organization?} 
\end{itemize}
\textbf{How did you know about this organization?}

\item \textbf{I advertised the opportunity directly on a mailing list, crowdworker platform, on the wall or train, on a website, etc.} This corresponds to the ``Infrastructural" code.

\begin{itemize}
 \item \textbf{On what mode of communication did you advertise?}
 \end{itemize}

\item \textbf{I leveraged existing personal networks as a recruiting strategy.} 
This corresponds with the ``Personal Networks" code.

\begin{itemize}
\item
 \textbf{Describe how the personal network was used in recruiting and what the personal connection was.} 
 \end{itemize}

\item \textbf{I met the participant at an event, such as a conference, a civic advocacy meeting, etc.} This corresponds to the ``Event" code.

\begin{itemize}
\item
 \textbf{Which event?} 
\end{itemize}

\end{enumerate}

\begin{itemize}
\item
\textbf{What important lessons did you learn from recruiting this stakeholder?} This question is added to encourage reflection on the recruitment process and to share any special learnings about the recruitment strategy. 
\end{itemize}

\begin{itemize}
\item \textbf{What did the call for participation look like?} This question is included to improve transparency of the recruitment process and provide concrete examples of how to design a call for participation for similar case studies.
\end{itemize}

\section{Tables}\label{sec:tables}
We next include tables with the results of coding our corpus (Tables \ref{table:recruiting_strategy},  \ref{table:who_initiated}, \ref{table:whole_story}, \ref{table:stakeholder_groups}, \ref{table:empowerment}, \ref{table:specificity}). These results inspired the questions included in ParticipAIte.

\begin{table*}
    \begin{tabular}{|p{0.2\linewidth}|p{0.4\linewidth}|p{0.3\linewidth}|}
            \hline
            Recruiting Strategy & Subcodes & Case studies \\
            \hline
            Organizational Recruitment (22) &
            Institutions (Universities); Online Communities; Crowdsourcing platforms; Organizations (community advocacy organizations); Community research groups; Monitoring Boards; Advisory Boards; Non governmental organizations; Social service providers and programs; Worker advocacy organizations; Employers and corporations
            & \cite{young_toward_2019,suresh_towards_2022,lee_webuildai_2019,zhu_value-sensitive_2018,newman_kudu_nodate,cheng_soliciting_2021,yu_keeping_2020,shi_improving_2020,queerinai_queer_2023,delgado_uncommon_2022,frauenberger_pursuit_2015,imran_rapid_2020,eitzel_assessing_2021,sendak__2020,service_goat_2019,wilder_clinical_2021,kuo_understanding_2023,flanigan_fair_2021,calacci_bargaining_2022,katell_toward_2020,irani_turkopticon_2013,marian_algorithmic_2023} 
            \\
            \hline
            Infrastructural Recruitment (17) &
            Social Media; Social media groups; Advertising; Radio; Websites; Talk Pages; Forums; Crowdsourcing platforms; News Articles; Publications; Loudspeaker announcements; Posters; Slack channels; Email listservs; Press Coverage; Bicycle shops; Bus stops
            &
            \cite{nekoto_participatory_2020,zhu_value-sensitive_2018,freedman_adapting_2020,newman_kudu_nodate,cheng_soliciting_2021,yu_keeping_2020,puussaar_sensemystreet_2022,shen_model_2022,halfaker_ores_2020,li_participation_2023,awad_moral_2018,bakker_fine-tuning_2022,deng_imagenet_2009,swanson_snapshot_2015,calacci_bargaining_2022,kelling_ebird_2012,irani_turkopticon_2013} \\
            \hline
            Personal Networks (10) & 
            Individual Networks; Volunteering; House to House knocking; Snowball Sampling; Direct contacts; Word-of-mouth; Cold email; Attending summer schools and programs
            & \cite{nekoto_participatory_2020,puussaar_sensemystreet_2022,shen_model_2022,halfaker_ores_2020,eitzel_assessing_2021,kuo_understanding_2023,calacci_bargaining_2022,kelling_ebird_2012,irani_turkopticon_2013, wilder_clinical_2021}  \\
            \hline
            Events (6) & 
            Professional conferences; Civic meetups; Trainings; Village Meetings; Exploratory Workshops; Workshops; Networking Events; Presentations
            &\cite{nekoto_participatory_2020,puussaar_sensemystreet_2022,queerinai_queer_2023,delgado_uncommon_2022,sendak__2020,irani_turkopticon_2013} \\
            \hline 
            Not Documented for at least one participant group & & \cite{finn_developing_2022, liao_enabling_2019, sinders_feminist_2020, imran_rapid_2020, bondi_spot_2018} \\ 
            \hline
    \end{tabular}
    \caption{What were the recruitment strategies?}
    \label{table:recruiting_strategy}
\end{table*}

\begin{table*}[!t]
\begin{minipage}{0.48\textwidth}
\centering

    \begin{tabular}{|p{0.2\linewidth}|p{0.75\linewidth}|}
            \hline
            Who initiated the recruitment strategy? & Case Studies \\
            \hline
             Researchers (24) & \cite{young_toward_2019,lee_webuildai_2019,zhu_value-sensitive_2018,freedman_adapting_2020,newman_kudu_nodate,cheng_soliciting_2021,yu_keeping_2020,puussaar_sensemystreet_2022,liao_enabling_2019,shen_model_2022,frauenberger_pursuit_2015,eitzel_assessing_2021,service_goat_2019,wilder_clinical_2021,awad_moral_2018, bakker_fine-tuning_2022,deng_imagenet_2009,swanson_snapshot_2015,kuo_understanding_2023,katell_toward_2020,marian_algorithmic_2023,irani_turkopticon_2013, sinders_feminist_2020, kelling_ebird_2012}  \\
            \hline
            Participants (5) & \cite{delgado_uncommon_2022,eitzel_assessing_2021,service_goat_2019,calacci_bargaining_2022, sendak__2020} \\
            \hline
            No clear delineation between the two roles (4) & \cite{nekoto_participatory_2020,queerinai_queer_2023,li_participation_2023,halfaker_ores_2020}  \\
            \hline
            Unclear for at least one participant group (6) & \cite{finn_developing_2022,suresh_towards_2022,shi_improving_2020,imran_rapid_2020,flanigan_fair_2021,calacci_bargaining_2022} \\
            \hline
    \end{tabular}
    \caption{Who initiated the recruitment?}
    \label{table:who_initiated}
    \end{minipage}
\hfill
\begin{minipage}{0.48\textwidth}
\centering
    \begin{tabular}{|p{0.3\linewidth}|p{0.65\linewidth}|}
            \hline
            Who were the different stakeholders involved? & Papers \\
            \hline
             Clients (9) & \cite{nekoto_participatory_2020,suresh_towards_2022,lee_webuildai_2019,zhu_value-sensitive_2018,shi_improving_2020,queerinai_queer_2023,halfaker_ores_2020,sendak__2020,service_goat_2019}\\
            \hline
            Collaborators (17) & \cite{nekoto_participatory_2020,finn_developing_2022,suresh_towards_2022,lee_webuildai_2019,zhu_value-sensitive_2018,newman_kudu_nodate,yu_keeping_2020,shi_improving_2020,shen_model_2022,queerinai_queer_2023,delgado_uncommon_2022,imran_rapid_2020,sendak__2020,flanigan_fair_2021,bondi_spot_2018, eitzel_assessing_2021, wilder_clinical_2021} \\
            \hline
            Decision Users (24) & \cite{nekoto_participatory_2020,finn_developing_2022,suresh_towards_2022,lee_webuildai_2019,zhu_value-sensitive_2018,cheng_soliciting_2021,newman_kudu_nodate,yu_keeping_2020,puussaar_sensemystreet_2022,shi_improving_2020,queerinai_queer_2023,delgado_uncommon_2022,shen_model_2022,halfaker_ores_2020,eitzel_assessing_2021,frauenberger_pursuit_2015,service_goat_2019,li_participation_2023,wilder_clinical_2021,flanigan_fair_2021,calacci_bargaining_2022,bondi_spot_2018,marian_algorithmic_2023, sendak__2020, wilder_clinical_2021}  \\
            \hline
            Data Creators (23) & \cite{nekoto_participatory_2020,finn_developing_2022,suresh_towards_2022,lee_webuildai_2019,zhu_value-sensitive_2018,freedman_adapting_2020,newman_kudu_nodate,yu_keeping_2020,puussaar_sensemystreet_2022,queerinai_queer_2023,liao_enabling_2019,delgado_uncommon_2022,sinders_feminist_2020,halfaker_ores_2020,imran_rapid_2020,awad_moral_2018,bakker_fine-tuning_2022,deng_imagenet_2009,swanson_snapshot_2015,calacci_bargaining_2022,kelling_ebird_2012,irani_turkopticon_2013,katell_toward_2020}. \\
            \hline Decision Subjects (14) & \cite{young_toward_2019,marian_algorithmic_2023,katell_toward_2020,calacci_bargaining_2022,kuo_understanding_2023,shen_model_2022,li_participation_2023,wilder_clinical_2021,service_goat_2019,queerinai_queer_2023,cheng_soliciting_2021,zhu_value-sensitive_2018, lee_webuildai_2019,frauenberger_pursuit_2015} \\
            \hline Regulators \& Governance (6) & \cite{finn_developing_2022,young_toward_2019,lee_webuildai_2019,queerinai_queer_2023,eitzel_assessing_2021,sendak__2020} \\ \hline
    \end{tabular}
    \caption{Who were the different stakeholders involved?}
    \label{table:stakeholder_groups}
\end{minipage}
\end{table*}

\begin{table}[t]
\begin{tabular}{|p{0.2\textwidth}|p{0.25\textwidth}|}
\hline
\textbf{Value} & \textbf{Case Studies} \\
\hline
All stakeholders have recruitment strategies documented & 
\cite{nekoto_participatory_2020,young_toward_2019,lee_webuildai_2019,zhu_value-sensitive_2018,freedman_adapting_2020,cheng_soliciting_2021,yu_keeping_2020,queerinai_queer_2023,shen_model_2022,halfaker_ores_2020,sendak__2020,service_goat_2019,li_participation_2023,awad_moral_2018,bakker_fine-tuning_2022,kuo_understanding_2023,irani_turkopticon_2013}
\\ \hline

Not all stakeholders have recruitment strategies documented & 
\cite{finn_developing_2022,suresh_towards_2022,newman_kudu_nodate,puussaar_sensemystreet_2022,shi_improving_2020,liao_enabling_2019,delgado_uncommon_2022,sinders_feminist_2020,frauenberger_pursuit_2015,imran_rapid_2020,eitzel_assessing_2021,wilder_clinical_2021,deng_imagenet_2009,swanson_snapshot_2015,flanigan_fair_2021,calacci_bargaining_2022,kelling_ebird_2012,katell_toward_2020,bondi_spot_2018,marian_algorithmic_2023}
\\ \hline

\end{tabular}

\caption{Are all the stakeholder recruitment strategy documented?}
\label{table:whole_story}

\end{table}

\begin{table}
\begin{tabular}{|p{0.25\textwidth}|p{0.25\textwidth}|}
\hline
\textbf{Empowerment Code} & \textbf{Papers} \\ \hline

Collaborate & \cite{katell_toward_2020,wood_ebird_2011,calacci_bargaining_2022,wilder_clinical_2021,halfaker_ores_2020,delgado_uncommon_2022,shi_improving_2020,queerinai_queer_2023,finn_developing_2022,lee_webuildai_2019,suresh_towards_2022} \\ \hline
Consult & \cite{awad_moral_2018,swanson_snapshot_2015,bakker_fine-tuning_2022,imran_rapid_2020,liao_enabling_2019,service_goat_2019,newman_kudu_nodate,freedman_adapting_2020,cheng_soliciting_2021} \\ \hline
Include & \cite{marian_algorithmic_2023,kuo_understanding_2023,bakker_fine-tuning_2022,irani_turkopticon_2013,kelling_ebird_2012,calacci_bargaining_2022,sendak__2020,sinders_feminist_2020,imran_rapid_2020,shen_model_2022,frauenberger_pursuit_2015,service_goat_2019,yu_keeping_2020,queerinai_queer_2023,zhu_value-sensitive_2018,young_toward_2019} \\ \hline
Other (Not documented or Difficult to Apply the scale) & \cite{bondi_spot_2018,marian_algorithmic_2023,kuo_understanding_2023,swanson_snapshot_2015,flanigan_fair_2021,deng_imagenet_2009,calacci_bargaining_2022,wilder_clinical_2021} \\ \hline
Own & 
\cite{kuo_understanding_2023,calacci_bargaining_2022,eitzel_assessing_2021,sinders_feminist_2020,wilder_clinical_2021,li_participation_2023,delgado_uncommon_2022,puussaar_sensemystreet_2022,suresh_towards_2022,nekoto_participatory_2020}

\\ \hline
\end{tabular}
\caption{Delgado's Empowerment Code}
\label{table:empowerment}

\end{table}

\begin{table}[t]
    \begin{tabular}{|l|p{0.25\textwidth}|}
    \hline
    \textbf{Specificity} & \textbf{Papers} \\
    \hline
    By Name & 
    \cite{nekoto_participatory_2020,suresh_towards_2022,lee_webuildai_2019,freedman_adapting_2020,liao_enabling_2019,service_goat_2019,eitzel_assessing_2021,sendak__2020,imran_rapid_2020,awad_moral_2018,deng_imagenet_2009,swanson_snapshot_2015,flanigan_fair_2021,calacci_bargaining_2022,kelling_ebird_2012,katell_toward_2020,bondi_spot_2018,marian_algorithmic_2023} (18)
    \\ \hline
    
    By Kind & 
    \cite{finn_developing_2022,young_toward_2019,newman_kudu_nodate,cheng_soliciting_2021,sinders_feminist_2020,frauenberger_pursuit_2015,wilder_clinical_2021,bakker_fine-tuning_2022,kuo_understanding_2023} (9)
    \\ \hline
    
    By Kind and Name & 
    \cite{zhu_value-sensitive_2018,yu_keeping_2020,puussaar_sensemystreet_2022,shi_improving_2020,queerinai_queer_2023,delgado_uncommon_2022,shen_model_2022,halfaker_ores_2020,li_participation_2023,irani_turkopticon_2013} (10)
    \\ \hline
    
\end{tabular}

\caption{How specific are the institutions, platforms, and organizations?}
\label{table:specificity}
\end{table}

\section{Corpus Tables}
We also present statistics related to the corpus. Table \ref{tab:country} sorts projects in our corpus by represented countries. Table \ref{tab:domains} sorts projects by domains. Table \ref{tab:entry_point} sorts projects by the participation entry point defined by \cite{corbett_power_2023}. 

\section{Corpus Links}

Finally, we include the 37 case studies in our corpus in table format with links. Links are verified at the time of publication in 2025. If copying a link from a PDF, please be sure to include underscores (\_).

\begin{table}[ht!]
\centering
\begin{tabular}{|l|c|}
\hline
\textbf{Country} & \textbf{Number} \\ \hline
Botswana        & 1 \\ \hline
Ethiopia        & 1 \\ \hline
Global          & 10 \\ \hline
Netherlands     & 1 \\ \hline
New Zealand     & 1 \\ \hline
South Africa    & 1 \\ \hline
Tanzania        & 1 \\ \hline
Uganda          & 1 \\ \hline
United Kingdom  & 6 \\ \hline
United States   & 17 \\ \hline
Uruguay         & 1 \\ \hline
Zimbabwe        & 2 \\ \hline
\end{tabular}
\caption{Table of Numbers by Country}
\label{tab:country}
\end{table}

\begin{table}[ht!]
    \centering
    \begin{tabular}{|l|c|}
        \hline
        \textbf{Category} & \textbf{Count} \\ \hline
        Low resourced languages & 2 \\ \hline
        Tech Policy & 2 \\ \hline
        Social Justice & 1 \\ \hline
        Humanitarian & 4 \\ \hline
        Wikipedia & 3 \\ \hline
        Health & 3 \\ \hline
        Agriculture & 2 \\ \hline
        Public Services & 2 \\ \hline
        Civics & 2 \\ \hline
        LGBTQIA+ & 2 \\ \hline
        AI Governance & 3 \\ \hline
        Law & 1 \\ \hline
        Education & 2 \\ \hline
        Social Media & 1 \\ \hline
        Computer Vision & 1 \\ \hline
        Conservation & 3 \\ \hline
        Housing & 1 \\ \hline
        Labor & 2 \\ \hline
    \end{tabular}
    \caption{Domains}
    \label{tab:domains}
\end{table}

\begin{table}[ht!]
    \centering
    \begin{tabular}{|l|c|}
        \hline
        \textbf{Category} & \textbf{Count} \\ \hline
        All points & 11 \\ \hline
        Dataset Development & 10 \\ \hline
        Deployment and Monitoring & 11 \\ \hline
        Problem Formulation & 8 \\ \hline
        Model Design and Training & 5 \\ \hline
        N/A & 1 \\ \hline
    \end{tabular}
    \caption{Participation Entry Point}
    \label{tab:entry_point}
\end{table}

\onecolumn
\begin{longtable}{|p{0.3\linewidth}|p{0.7\linewidth}|}
\caption{Citations of Corpus and Corresponding Links} \\
\hline
\textbf{Citation} & \textbf{Link} \\
\hline
\endhead
\citet{awad_moral_2018} & \url{https://www.nature.com/articles/s41586-018-0637-6} \\
\hline
\citet{bakker_fine-tuning_2022} & \url{https://proceedings.neurips.cc/paper_files/paper/2022/file/f978c8f3b5f399cae464e85f72e28503-Paper-Conference.pdf} \\
\hline
\citet{bondi_spot_2018} & \url{https://www.cais.usc.edu/wp-content/uploads/2017/11/spot-camera-ready.pdf} \\
\hline
\citet{calacci_bargaining_2022} & \url{https://dl.acm.org/doi/10.1145/3570601} \\
\hline
\citet{cheng_soliciting_2021} & \url{https://dl.acm.org/doi/fullHtml/10.1145/3411764.3445308} \\
\hline
\citet{delgado_uncommon_2022} & \url{https://dl.acm.org/doi/10.1145/3512898} \\
\hline
\citet{deng_imagenet_2009} & \url{https://ieeexplore.ieee.org/stamp/stamp.jsp?tp=&arnumber=5206848} \\
\hline
\citet{eitzel_assessing_2021} & \url{https://theoryandpractice.citizenscienceassociation.org/articles/10.5334/cstp.339} \\
\hline
\citet{finn_developing_2022} & \url{https://aclanthology.org/2022.computel-1.12.pdf} \\
\hline
\citet{flanigan_fair_2021} & \url{https://www.nature.com/articles/s41586-021-03788-6} \\
\hline
\citet{frauenberger_pursuit_2015} & \url{https://www.sciencedirect.com/science/article/pii/S1071581914001232} \\
\hline
\citet{freedman_adapting_2020} & \url{https://arxiv.org/abs/2005.09755} \\
\hline
\citet{halfaker_ores_2020} & \url{https://arxiv.org/pdf/1909.05189} \\
\hline
\citet{imran_rapid_2020} & \url{https://arxiv.org/abs/2004.06675} \\
\hline
\citet{irani_turkopticon_2013} & \url{https://escholarship.org/content/qt10c125z3/qt10c125z3_noSplash_17ace960c310470e904a7512eccf6ee2.pdf} \\
\hline
\citet{katell_toward_2020} & \url{https://dl.acm.org/doi/10.1145/3351095.3372874} \\
\hline
\citet{kelling_ebird_2012} \footnote{Same case study as \cite{wood_ebird_2011}} & \url{https://ojs.aaai.org/index.php/AAAI/article/view/18963} \\
\hline
\citet{kuo_understanding_2023} & \url{https://dl.acm.org/doi/10.1145/3544548.3580882} \\
\hline
\citet{lee_webuildai_2019} & \url{https://dl.acm.org/doi/10.1145/3359283} \\
\hline
\citet{li_participation_2023} & \url{https://dl.acm.org/doi/10.1145/3544548.3582074} \\
\hline
\citet{liao_enabling_2019} & \url{https://arxiv.org/abs/1912.07381} \\
\hline
\citet{marian_algorithmic_2023} & \url{https://dl.acm.org/doi/pdf/10.1145/3593013.3594009} \\
\hline
\citet{nekoto_participatory_2020} & \url{https://arxiv.org/abs/2010.02353} \\
\hline
\citet{newman_kudu_nodate} & \url{https://ai4sibook.org/kudu/} \\
\hline
\citet{puussaar_sensemystreet_2022} & \url{https://dl.acm.org/doi/abs/10.1145/3555215} \\
\hline
\citet{queerinai_queer_2023} & \url{https://dspace.mit.edu/handle/1721.1/151097} \\
\hline
\citet{sendak__2020} & \url{https://dl.acm.org/doi/10.1145/3351095.3372827} \\
\hline
\citet{service_goat_2019} & \url{https://medium.com/unhcr-innovation-service/a-goat-story-3ed6bdd2b237} \\
\hline
\citet{shen_model_2022} & \url{https://dl.acm.org/doi/pdf/10.1145/3531146.3533110} \\
\hline
\citet{shi_improving_2020} & \url{https://ai4sibook.org/wp-content/uploads/2022/08/Food-Rescue.pdf} \\
\hline
\citet{sinders_feminist_2020} & \url{https://carolinesinders.com/wp-content/uploads/2020/05/Feminist-Data-Set-Final-Draft-2020-0526.pdf} \\
\hline
\citet{suresh_towards_2022} & \url{https://dl.acm.org/doi/pdf/10.1145/3531146.3533132} \\
\hline
\citet{swanson_snapshot_2015} & \url{https://www.nature.com/articles/sdata201526} \\
\hline
\citet{wilder_clinical_2021} & \url{https://ai4sibook.org/hiv/} \\
\hline
\citet{wood_ebird_2011} \footnote{Same case study as \cite{kelling_ebird_2012}} & \url{https://journals.plos.org/plosbiology/article?id=10.1371/journal.pbio.1001220} \\
\hline
\citet{young_toward_2019} & \url{https://link.springer.com/article/10.1007/s10676-019-09497-z} \\
\hline
\citet{yu_keeping_2020} & \url{https://dl.acm.org/doi/10.1145/3357236.3395528} \\
\hline
\citet{zhu_value-sensitive_2018} & \url{https://dl.acm.org/doi/10.1145/3274463} \\
\hline
\end{longtable}

\end{document}